\documentclass[preprint]{aastex6}
\usepackage{epsfig}
\usepackage{graphicx}
\usepackage{dcolumn}
\usepackage{bm}
\usepackage{amsmath}

\shorttitle{Thermonuclear $^{17}$O($n$,$\gamma$)$^{18}$O reaction rate and ...}
\shortauthors{Li-Yong Zhang et al.}

\usepackage{xcolor}
\begin{document}


\title{Thermonuclear $^{17}$O($\lowercase{n}$,$\gamma$)$^{18}$O reaction rate and its astrophysical implications}


\author{Li-Yong Zhang\altaffilmark{1,2}, Jian-Jun He\altaffilmark{1,2}, Motohiko Kusakabe\altaffilmark{3}, Zhen-Yu He\altaffilmark{3}, Toshitaka Kajino\altaffilmark{3,4,5}}
\affil{\altaffilmark{1}Key Laboratory of Beam Technology of Ministry of Education, College of Nuclear Science and Technology, Beijing Normal University, Beijing 100875, China}
\affil{\altaffilmark{2}Beijing Radiation Center, Beijing 100875, China}
\affil{\altaffilmark{3}School of Physics, and International Research Center for Big-Bang Cosmology and Element Genesis, Beihang University, Beijing 100191, China}
\affil{\altaffilmark{4}Department of Astronomy, School of Science, the University of Tokyo, 7-3-1 Hongo, Bunkyo-ku, Tokyo, 113-0033, Japan}
\affil{\altaffilmark{5}National Astronomical Observatory of Japan 2-21-1 Osawa, Mitaka, Tokyo, 181-8588, Japan}

\email{Corresponding author emails: hejianjun@bnu.edu.cn, kusakabe@buaa.edu.cn, kajino@buaa.edu.cn}

\begin{abstract}
A new thermonuclear $^{17}$O($n$,$\gamma$)$^{18}$O rate is derived based on a complete calculation of the direct-capture (DC) and resonant-capture contributions, for a
temperature region up to 2 GK of astrophysical interest. We have firstly calculated the DC and subthreshold contributions in the energy region up to 1 MeV, and estimated the
associated uncertainties by a Monte-Carlo approach. It shows that the present rate is remarkably larger than that adopted in the JINA REACLIB in the temperature region of
0.01 $\sim$ 2 GK, by up to a factor of $\sim$80. The astrophysical impacts of our rate have been examined in both $s$-process and $r$-process models. In our main
$s$-process model which simulates flash-driven convective mixing in metal deficient asymptotic giant branch stars, both $^{18}$O and $^{19}$F abundances in interpulse
phases are enhanced dramatically by factors of $\sim 20$--$40$ due to the new larger $^{17}$O($n$,$\gamma$)$^{18}$O rate. It shows, however, that this reaction hardly
affects the weak $s$-process in massive stars since the $^{17}$O abundance never becomes significantly large in the massive stars. For the $r$-process nucleosynthesis,
we have studied impacts of our rate in both the collapsar and neutron burst models, and found that the effect can be neglected, although an interesting ``loophole" effect is
found owing to the enhanced new rate, which significantly changes the final nuclear abundances if fission recycling is not involved in the model, however, these significant
differences are almost completely washed out if the fission recycling is considered.

\end{abstract}


\section{Introduction}
\label{sec1}
In He shell flash models, both $^{16}$O and $^{17}$O are on the pathway to heavier elements in the nucleosynthesis process, and the nuclear reaction $^{17}$O($n$,$\gamma$)$^{18}$O
is expected to play a crucial role. \citet{yam10} studied this reaction and demonstrated how the $^{17}$O($n$,$\gamma$)$^{18}$O rate changes the synthesis paths to resultant
heavier elements. Because there were no experimental data available for this ($n$,$\gamma$) cross section at that time, they assumed two limits for the rate at $kT$ = 30 keV, i.e.,
the lower limit (Data-A) of $N_\mathrm{A}$$\langle \sigma v \rangle$ = 1.3$\times$10$^3$ cm$^3$s$^{-1}$mol$^{-1}$ and the upper limit (Data-B) of
$N_\mathrm{A}$$\langle \sigma v \rangle$ = 1.5$\times$10$^6$ cm$^3$s$^{-1}$mol$^{-1}$, and also assumed the constant reaction rate w.r.t. temperature in their calculations.
It was shown that the abundances predicted for 5 $<Z<$ 16 elements, the number abundance of neutron, as well as the synthesis pathway from C to Ne, can be largely affected
by such rate changes. In fact, this lower limit corresponds to the $^{18}$O($n$,$\gamma$)$^{19}$O rate at $kT$ = 30 keV, and the upper limit corresponds to that compiled in the
Brussels Nuclear Library for Astrophysics Applications (BRUSLIB)\footnote{http://www.astro.ulb.ac.be/bruslib/}~\citep{gor02,arn06,xu13}. Later on,~\citet{nis09} studied the role of
$^{17}$O neutron capture reactions in the metal-free and extremely metal-poor asymptotic giant branch (AGB) stars, by assuming the $^{17}$O($n$,$\gamma$)$^{18}$O rate equal to the
$^{16}$O($n$,$\gamma$)$^{17}$O rate. The orders of magnitude uncertainty previously assumed for the $^{17}$O($n$,$\gamma$)$^{18}$O rate is indeed too large to constrain the
stellar evolution models.

Massive stars with initial mass $M>8 M_\sun$ experience the weak slow neutron capture process ($s$-process) during the He- and C-burning stages~\citep{lam77,rai91,rai93,the07}.
$^{16}$O is one of the most important neutron poisons in the weak $s$-process~\citep[e.g.,][]{moh16}, and $^{17}$O nuclei are produced via the reaction
$^{16}$O($n$,$\gamma$)$^{17}$O. This reaction significantly consumes neutrons produced via $^{22}$Ne($\alpha$,$n$)$^{25}$Mg, and abundances of light nuclei C to Mg evolve
sensitively to the reaction rate in the weak $s$-process~\citep{HeM20}. The subsequent reaction $^{17}$O($n$,$\gamma$)$^{18}$O on the product $^{17}$O is also involved in a reaction
network of the light nuclei that determine the neutron budget during the weak $s$-process. In addition, this neutron capture reaction is also involved {bf in the main $s$-process nucleosynthesis in AGB stars~\citep{koeppeler11} and} rapid neutron capture
process ($r$-process) nucleosynthesis~\citep{rau94}. Therefore, a more reliable rate for the $^{17}$O($n$,$\gamma$)$^{18}$O reaction is highly desirable for the nuclear
astrophysics community.

About twenty years ago, ~\citet{koe91} derived an analytic reaction rate formula for the $^{17}$O($n$,$\gamma$)$^{18}$O reaction. It was based on a thermal ($n$,$\gamma$) cross
section of $\sigma_\mathrm{th}$ = (0.538$\pm$0.065) mb and two known resonances at $E_\mathrm{R}$ = 0.169, 0.238 MeV (in the center-of-mass frame). This experimental thermal
$\sigma_\mathrm{th}$ value was originally reported by~\citet{lon79} in a Conference proceedings and later compiled into a book by~\citet{mug81}. This reaction rate was utilized in
the study of heavy element production in inhomogeneous cosmologies~\citep{rau94}, and referred to as \emph{bb92} in the JINA
REACLIB\footnote{http://groups.nscl.msu.edu/jina/reaclib/db}~\citep{cyb10}. Recently, a more precise value of $\sigma_\mathrm{th}$ = (0.67$\pm$0.07) mb has been obtained
experimentally~\citep{fir16}. \citet{pri10} calculated the Maxwellian-averaged cross sections (MACSs) and astrophysical reaction rates using the evaluated nuclear reaction data
libraries, where the MACSs and rates of $^{17}$O($n$,$\gamma$)$^{18}$O were listed in tables at 15 data points in an energy/temperature range of $kT$ = 1 $\sim$ 1000 keV, with
ENDF/B-VII.0~\citep{cha06} \& B-VI.8~\citep{bvi8} and JEFF-3.1~\citep{jac02} libraries, respectively. In fact, these three libraries gave the exactly same results. Furthermore, in
BRUSLIB, the thermonuclear reaction rates for nuclei with $Z$ = 8--109 were estimated using the Hauser-Feshbach statistical model. However, it has been shown that this statistical
model is highly applicable to heavy nuclei which have large nuclear level densities at excitation energies of a few MeV. Actually, overestimations in the BRUSLIB theoretical values can
be seen when the mass number decreases below $A<$ 50. For instance, the $^{17}$O($n$,$\gamma$)$^{18}$O rates estimated in BRUSLIB is about two orders of magnitude larger
than those rates mentioned above.

In this work, we have obtained the thermonuclear $^{17}$O($n$,$\gamma$)$^{18}$O rate based on theoretical calculation of the direct capture (DC) as well as the resonance
contributions, for a temperature up to $T_9$ = 2 of general astrophysical interest. Three rates based on the recent evaluated nuclear reaction data libraries are also calculated and
compared. It shows significantly large discrepancies between different rates. For the higher temperature region of $T_9$ = 2 $\sim$ 10, the $^{17}$O($n$,$\gamma$)$^{18}$O rate
is calculated by using the high-energy evaluated data in ENDF/B-VIII with a normalized factor of 1.45. The impacts of our new rate have been investigated in stellar $s$-process and $r$-process models, and many interesting characters have been revealed for the present rate.

\section{Present rates}
\label{sec2}
To calculate the thermonuclear reaction rate, we have adopted the method described by~\citet{pri10}, which will be introduced simply as follows. The MACS is defined as
\begin{eqnarray}
\sigma^\mathrm{MACS}(kT) = \frac{\langle \sigma v \rangle}{v_T},
\label{eq1}
\end{eqnarray}
where $v$ is the relative velocity of neutron with respect to a target nuclide, and $v_T$ is the mean thermal velocity given by
\begin{eqnarray}
v_T = \sqrt{\frac{2kT}{\mu}},
\label{eq2}
\end{eqnarray}
where $\mu$ is the reduced mass of the target nucleus and neutron. Here, nuclear masses for neutron and $^{17}$O are $m_n$ = 1.008665~u and
$m\mathrm{(^{17}O)}$ = 16.994743~u (with u in the atomic mass unit), based on AME2016~\citep{wan17}, respectively. The MACS can be calculated as
\begin{eqnarray}
\sigma^\mathrm{MACS}(kT) = \frac{2}{\sqrt{\pi}}\times\frac{1}{(kT)^2} \int^{\infty}_{0} \sigma(E)E\mathrm{exp}\left(-\frac{E}{kT}\right)dE,
\label{eq3}
\end{eqnarray}
where $E$ is the energy of relative motion of the neutron with respect to the target, i.e., the center-of-mass energy $E_\mathrm{c.m.}$. It should be noted that a format of neutron
energy $E_n$ $v.s.$ $\sigma$ is adopted in all the evaluated nuclear reaction data libraries, and hence one should be careful in using Eq.~\ref{eq3}. The astrophysical reaction rate,
$R$, is defined as $R=N_A \langle \sigma v \rangle$, with $N_A$ the Avogadro number. Finally, the rate can be calculated by
\begin{eqnarray}
R(T_9)=10^{-24}N_A\sigma^\mathrm{MACS}(kT)v_T,
\label{eq4}
\end{eqnarray}
with $R$ in units of cm$^3$s$^{-1}$mol$^{-1}$, $v_T$ in units of cm/s, and $\sigma^\mathrm{MACS}$ in units of barn. The temperature $kT$ in units of energy (e.g., MeV) is related
to that in Kelvin as $T_9$ = 11.605 $kT$ (here, $T_9$ = 1 means $T$ = 10$^9$ K). Therefore, once we obtain the cross section data in the energy region of astrophysical interest, the
rates can be numerically calculated by using the equations introduced above.

In this work, we have calculated the cross section of $^{17}$O($n$,$\gamma$)$^{18}$O reaction in an $E_\mathrm{c.m.}$ energy region up to 1 MeV, which are sufficient to account for
thermonuclear reaction rate up to a temperature of $T_9$ = 2. The total neutron capture rate is the sum of resonant- and direct-capture (DC) on the ground state and thermally excited
states in target nucleus, weighted with their individual population factors~\citep{rol88,lam16}. In the present case, such a thermal excitation effect can be neglected for the temperature
below $T_9$ = 2 because the first-excited state in target nucleus of $^{17}$O is quite high ($E_x$ = 870.73 keV). Therefore, we calculate here only the contributions of the ground-state
capture.

\subsection{DC contribution}
\label{sec2.1}
We have calculated the DC cross sections of $^{17}$O($n$,$\gamma$)$^{18}$O with a RADCAP code~\citep{ber03}. The Woods-Saxon nuclear potential (central + spin orbit) and a
Coulomb potential of uniform-charge distribution were utilized in the calculation. The nuclear central potential $V_0$ was determined by matching the bound-state energies. The
spectroscopic factors ($C^2S$) and the orbital ($nlj$) information were taken from~\citet{li76}. The optical-potential parameters adopted from~\citet{hua10} are:
$R_0=R_\mathrm{s.o.}=R_C=r_0(1+A_T)^{1/3}$ (with $r_0$ = 1.25 fm, $A_T$ = 17 for the present $^{17}$O case), $a_0=a_\mathrm{s.o.}=0.65$~fm, with a spin-orbit potential
depth of $V_\mathrm{s.o.}=-10$ MeV. Here, $R_0$, $R_\mathrm{s.o.}$, and $R_C$ are the radii of central potential, the spin-orbit potential and the Coulomb potential, respectively;
$a_0$ and $a_\mathrm{s.o.}$ are the corresponding diffuseness parameters in the central and spin-orbit potentials, respectively. Similarly to~\citet{hjj14}, the parameter ($r_0$, $a$
and $V_\mathrm{s.o.}$) dependence on cross sections has been studied, and the associated uncertainties are estimated to be no more than 50\% in the energy region studied.

\begin{figure}[t]
\begin{center}
\includegraphics[width=0.6\textwidth]{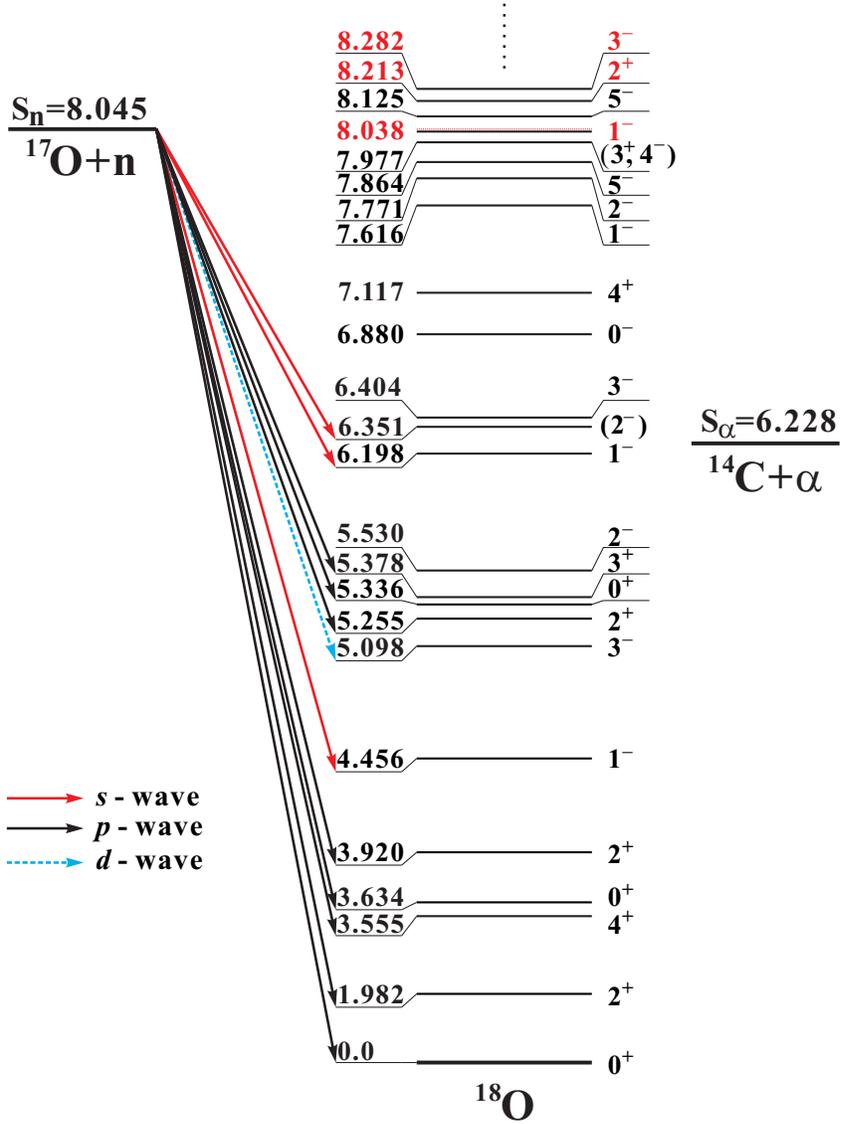}
\caption{\label{fig1} Level scheme of the $^{17}$O($n$,$\gamma$)$^{18}$O reaction. Three $s$-wave, eight $p$-wave and one $d$-wave captures are indicated by different colored
arrows; two resonances at $E_x$ = 8.213 and 8.282 MeV (i.e., $E_\mathrm{R}$ = 168 and 237 keV), and one subthreshold state at $E_x$ = 8.038 MeV (i.e., $E_\mathrm{R}$ = $-$7.6 keV),
are marked in red font. The location of $\alpha$ threshold is also shown. The data are adopted from~\citet{til95}.}
\end{center}
\end{figure}

\begin{figure}[t]
\begin{center}
\includegraphics[width=0.6\textwidth]{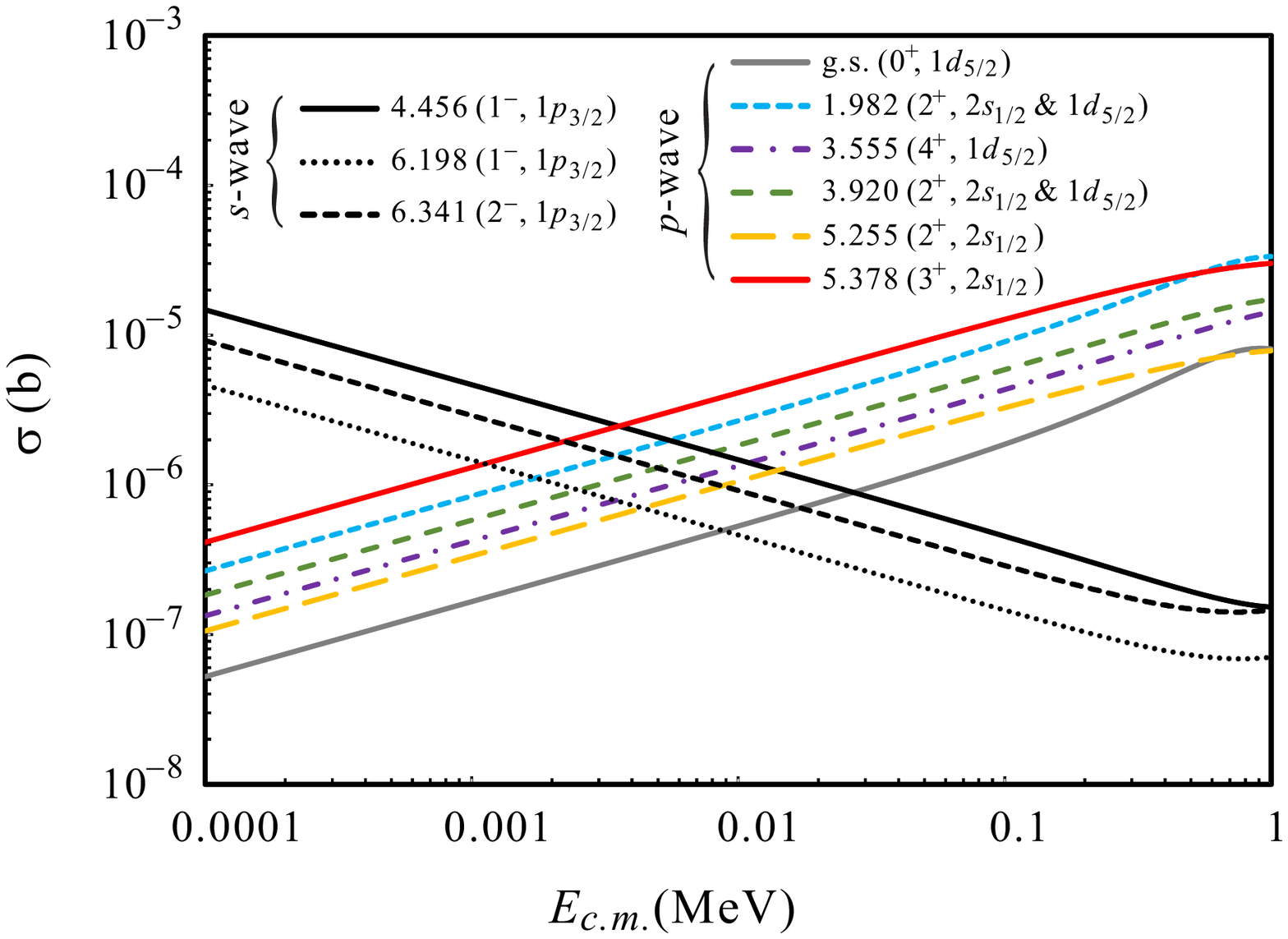}
\caption{\label{fig2} Major DC contributions calculated for the $^{17}$O($n$,$\gamma$)$^{18}$O reaction by a RADCAP code. Thereinto, $s$-wave and $p$-wave are categorized.}
\end{center}
\end{figure}

Figure~\ref{fig1} shows the level scheme of the $^{17}$O($n$,$\gamma$)$^{18}$O reaction. We have calculated twelve DC captures into the ground state and the excited states up to
$E_x$ = 6.351 MeV, if $C^2S$ factors are available. These captures include three $s$-wave, eight $p$-wave and one $d$-wave, which are indicated by three different colored arrows.
The calculation results are shown in Fig.~\ref{fig2}, where the major contributions from three $s$- and six $p$-wave captures are shown separately. The remaining one $d$- and two
$p$-wave contributions are negligibly small and not shown in the figure. It can be seen that $s$- and $d$-wave captures contribute equivalently to the cross sections around 3 keV, but
$s$- and $d$-wave become dominate below and above this energy, respectively. It should be noted that there are no any $d$-wave calculations available in the literature. Therefore, we
have made the $d$-wave calculations for the first time, and found that its contribution can not be neglected.

\begin{figure}[t]
\begin{center}
\includegraphics[width=0.6\textwidth]{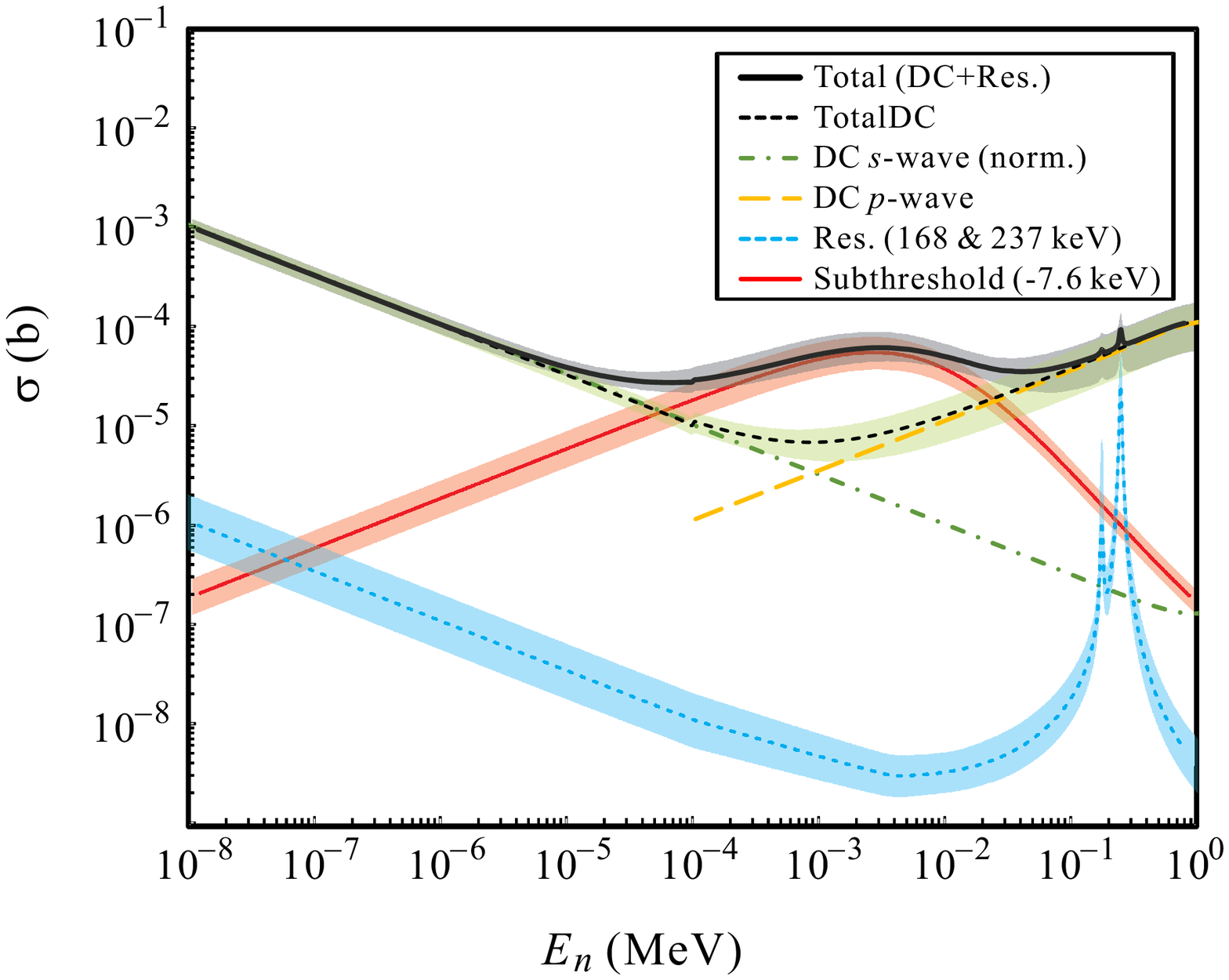}
\caption{\label{fig3} Major DC contributions calculated for the $^{17}$O($n$,$\gamma$)$^{18}$O reaction by a RADCAP code. Thereinto, $s$-wave and $p$-wave are categorized, and
the calculated $s$-wave contribution is normalized to the experimental value of $\sigma_\mathrm{th}$ = 0.67 mb. The corresponding uncertainties indicated as the error bands are
calculated by a Monte-Carlo method, see text for details.}
\end{center}
\end{figure}

It is well-known that the cross section of the $s$-wave neutron capture follows the 1/$v$ law. Accordingly, we predicted a thermal value of $\sigma_\mathrm{th}$ = 1.91 mb.
Compared to the recent experimental value of $\sigma_\mathrm{th}$ = (0.67$\pm$0.07) mb~\citep{fir16}, our predicted one is about a factor of 3 larger. This can be explained by the
possible uncertainties in the experimental $C^2S$ factors derived in~\citet{li76}, where no errors were quoted explicitly. All experimental $C^2S$ factors for three $s$-wave captured
states (at $E_x$ = 4.456, 6.198, 6.341 MeV) are 0.03, a very small quantity, and hence their errors may be large. The weak-coupling model predicted $C^2S$ = 2$\times$10$^{-4}$ and
0.013, respectively for the $E_x$ = 4.456 and 6.341 MeV states, which are about a factor of 3 or more smaller than the experimental ones. If we use weak-coupling predictions for these
three states, a value of $\sigma_\mathrm{th}$ = 0.58 mb is then obtained, which is consistent with the experimental result. Therefore, we normalized the $s$-wave results to the
experimental value of $\sigma_\mathrm{th}$ = 0.67 mb at the thermal energy. Figure~\ref{fig3} shows the summed $s$-wave (normalized) and $p$-wave contributions, as well as the
total DC contributions, respectively.

\subsection{Resonant contribution}
\label{sec2.2}
A general Breit-Wigner formula is utilized to calculate cross sections of the resonances~\citep{rol88,ili07},
\begin{eqnarray}
\sigma(E)=\pi \lambdabar^2 \omega \frac{\Gamma_n(E) \Gamma_\gamma(E)}{(E-E_\mathrm{R})^2+[\Gamma_\mathrm{tot}(E)/2]^2},
\label{eq5}
\end{eqnarray}
with the de Broglie wavelength $\lambda$ divided by 2$\pi$, i.e., $\lambdabar = \hbar/\sqrt{2\mu E}$, and the statistical factor
$\omega=(2J_\mathrm{R}+1)/[(2J_n+1)(2J_\mathrm{^{17}O}+1)]$. Here $J_\mathrm{R}$ is the spin of the resonance in the compound nucleus $^{18}$O, and the spins of projectile and
target nuclei are $J_n$ = 1/2, $J_\mathrm{^{17}O}$ = 5/2, respectively. All energies and widths are in the center-of-mass system, the partial widths $\Gamma_n(E)$ and
$\Gamma_\gamma(E)$, and thus the total width $\Gamma_\mathrm{tot}(E)$, are all energy dependent. In this work, one subthreshold state and two resonances have been considered,
and their resonance parameters are summarized in Table~\ref{tab1_par}.

The cross section at the resonant energy $E_\mathrm{R}$ is
$\sigma(E_\mathrm{R})=\sigma(E=E_\mathrm{R})=4\pi\lambdabar_\mathrm{R}^2\omega \Gamma_n\Gamma_\gamma/\Gamma_\mathrm{tot}^2$, with the total width
$\Gamma_\mathrm{R}=\Gamma_\mathrm{tot}(E=E_\mathrm{R})$, and thus the resonance Eq.~\ref{eq5} can be expressed as,
\begin{eqnarray}
\sigma(E)=\sigma(E_\mathrm{R})\frac{E_\mathrm{R}}{E}\frac{\Gamma_n(E)}{\Gamma_n(E_\mathrm{R})}\frac{\Gamma_\gamma(E)}{\Gamma_\gamma(E_\mathrm{R})} \frac{(\Gamma_\mathrm{R}/2)^2}{(E-E_\mathrm{R})^2+[\Gamma_\mathrm{tot}(E)/2]^2}.
\label{eq6}
\end{eqnarray}
Here, the total width includes three terms, i.e., $\Gamma_\mathrm{tot}(E)=\Gamma_\alpha(E)+\Gamma_n(E)+\Gamma_\gamma(E)$. The energy dependence of these three terms will
be explained below:
(1) for the dominant alpha width, the ratio can be calculated by
$\Gamma_\alpha(E)/\Gamma_\alpha(E_\mathrm{R}) =P_{\ell_\alpha}(E)/P_{\ell_\alpha}(E_\mathrm{R})$ with the penetration factor $P_{\ell_\alpha}$ for the $\alpha$ emission;
(2) for the low energy neutrons, the ratio can be calculated by $\Gamma_n(E)/\Gamma_n(E_\mathrm{R})=(E/E_\mathrm{R})^{\ell_n+1/2}$;
(3) for the gamma width, the ratio can be calculated by $\Gamma_\gamma(E)/\Gamma_\gamma(E_\mathrm{R}) = [(E + Q - E_f)/(E_\mathrm{R} + Q - E_f)]^{2L+1}$, where $Q$ is the
$Q$ value of ($n$,$\gamma$) reaction, and $E_f$ is the energy of the final state in the compound nucleus. $L$ is the multipolarity of the electromagnetic transition, $L$ = 1 for the major
transitions of the present $E1$ or $M1$ case. The calculations show that the energy dependence of $\Gamma_\alpha$ and $\Gamma_\gamma$ is not very large, while that of
$\Gamma_n$ is quite significant. Anyway, we have included all the energy dependence in the calculations. It is worthy of noting that we have derived the following equation that is used
for the subthreshold state (with a negative value of $E_\mathrm{R}$) calculation,
\begin{eqnarray}
\sigma(E)=\sigma(|E_\mathrm{R}|)\frac{|E_\mathrm{R}|}{E}\frac{\Gamma_n(E)}{\Gamma_n(|E_\mathrm{R}|)}\frac{\Gamma_\gamma(E)}{\Gamma_\gamma(|E_\mathrm{R}|)} \frac{(2E_\mathrm{R})^2+(\Gamma_\mathrm{R}/2)^2}{(E-E_\mathrm{R})^2+[\Gamma_\mathrm{tot}(E)/2]^2}.
\label{eq7}
\end{eqnarray}

\begin{table*}
\footnotesize
\caption{\label{tab1_par} Relevant nuclear levels in the compound nucleus $^{18}$O, including one subthreshold state and two resonances. Most of the parameters are adopted
from~\citet{sch93} unless otherwise specified. The reaction $Q$ value of $S_n$ = 8.045 MeV is taken from AME2016~\citep{wan17}.}
\begin{tabular}{ccccccc}
\hline
$E_x$ (keV)  & $E_\mathrm{R}$ (keV) & $J^\pi$ & $\Gamma_\mathrm{tot}$ (eV) & $\Gamma_\alpha$ (eV)$^a$ & $\Gamma_n$ (eV)$^b$ & $\Gamma_\gamma$ (eV)$^c$  \\
\hline
8037.8$^d$ & $-$7.6$\pm$0.7 & 1$^-$ & 2400$^e$      &	2300 [$\ell_\alpha$ = 1]         & 98 [$\ell_n$ = 1]          & 1.15$\pm$0.15 [E1]    \\
8213       & 168$\pm$4    & 2$^+$ & 1280$\pm$1000 &	1200$\pm$800 [$\ell_\alpha$ = 2] & 14$\pm$9 [$\ell_n$ = 0]    & 0.41$\pm$0.09 [E1+M1] \\
8282       & 237$\pm$3    & 3$^-$ & 8000$\pm$1000 &	6900$\pm$400 [$\ell_\alpha$ = 3] & 830$\pm$490 [$\ell_n$ = 1] & 0.49$\pm$0.13 [E1]    \\
\hline
\end{tabular}

$^a$: The transferred $\ell_\alpha$ value is listed in the square brackets;
$^b$: The transferred $\ell_n$ value is listed in the square brackets;
$^c$: The major $\gamma$-ray multipolarity is listed in the square brackets;
$^d$: The uncertainties in $\Gamma_\mathrm{tot}$, $\Gamma_\alpha$ and $\Gamma_n$ (eV) are estimated to be 20\% for this subthreshold state;
$^e$: An upper limit of 2500 eV was set by~\citet{ajz87,til95}.
\end{table*}

Figure~\ref{fig3} shows the contributions owing to the two resonances and one subthreshold state listed in Table~\ref{tab1_par}. We have made the first calculation for the subthreshold
state at $E_\mathrm{R}$ = $-$7.6 keV, and found that its contribution dominates the total cross sections in the energy range of $\sim$0.1--20 keV.

\subsection{Total cross section}
\label{sec2.3}
\begin{figure}[t]
\begin{center}
\includegraphics[width=0.6\textwidth]{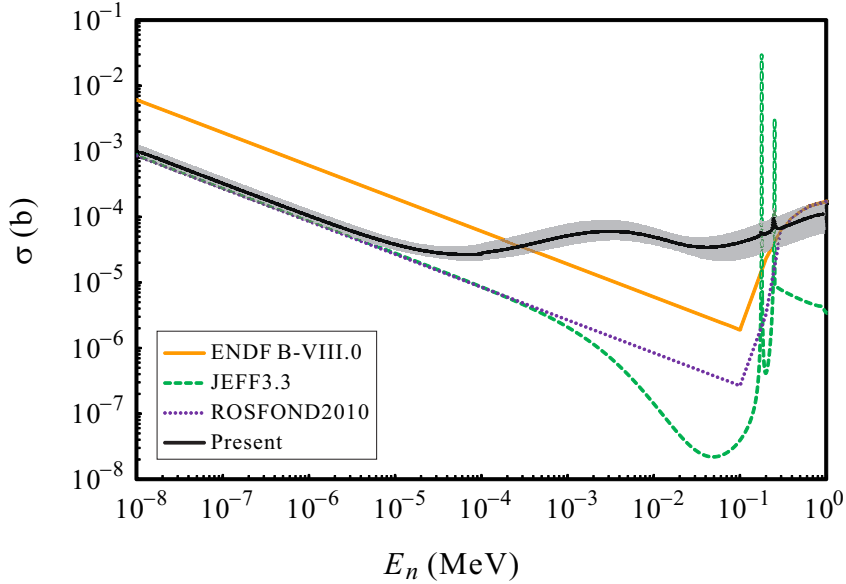}
\caption{\label{fig4} Cross sections of $^{17}$O($n$,$\gamma$)$^{18}$O calculated in this study as a function of incident neutron energy. The typical evaluated data are shown for
comparison. The uncertainties of the present cross section indicated as an error band are calculated by a Monte-Carlo method.}
\end{center}
\end{figure}

The total cross sections have been calculated by summing the contributions of DC and resonances (including subthreshold state) discussed above. The present total cross section is
shown as a black solid line in Fig.~\ref{fig3}, where respective contributions are illustrated separately. Figure~\ref{fig4} shows a comparison between present results with those evaluated
in various nuclear reaction data libraries. The typical libraries are, ENDF/B-VIII.0 in USA~\citep{bviii}, JEFF3.3 in Europe~\citep{jeff}, ROSFOND2010 in Russia~\citep{rosfond}, where
the data are available from the website: https://www.nndc.bnl.gov/exfor/endf00.jsp. We find that the evaluated $^{17}$O($n$,$\gamma$)$^{18}$O cross section data are almost the
same among various ENDF/B series libraries (i.e., B-V, B-VI, B-VII, B-VIII). A very large thermal value of $\sigma_\mathrm{th}$ = 3.83 mb was given in various ENDF/B libraries,
whose header information in the $^{17}$O($n$,$\gamma$)$^{18}$O data file stated as ``THERMAL CROSS SECTIONS N,GAMMA = 3.8 MB+/-20PC CALCULATED BY S.F.MUGHABGHAB
(MU78) USING METHOD OF LANE + LYNN (SEE DOCUMENT)''. The cited reference MU78 was a BNL report~\citep{mug78}, that was later published as ~\citet{mug79} in which no any
information about the $^{17}$O($n$,$\gamma$)$^{18}$O reaction was discussed. However, a value of $\sigma_\mathrm{th}$ = 0.38 mb was given in the book of~\citet{mug81}.
As explained by~\citet{pri12}, ``The ENDF/B-VII.1 thermal capture cross section, 3.8 mb, for $^{17}$O is attributed to a misquoted direct capture cross section calculated
by~\citet{mug81}''. Now, it's clear that all ENDF/B series gave a wrong $\sigma_\mathrm{th}$ value due to this misquotation. Therefore, the thermal value and associated header
information should be corrected in the future evaluations. ROSFOND2010 normalized the thermal cross section to the old experimental value of 0.538 mb~\citep{lon79} for the $s$-wave
dominated energy region below $E_n\sim$ 0.1 MeV, and adopted the high-energy data as in the ENDF/B libraries. The header of ROSFOND2010 data file for
$^{17}$O($n$,$\gamma$)$^{18}$O also included the same information of the thermal cross section as mentioned above, although it adopted an experimental thermal value. JEFF3.3,
which is based on a software system built round the nuclear model code TALYS~\citep{kon12}, gave a value of $\sigma_\mathrm{th}$ = 0.556 mb, and two resonant contributions were
also involved. It shows that the discrepancies between present results and the evaluated ones are remarkable.

The associated uncertainties in the present total cross sections and the reaction rates have been calculated by the Monte-Carlo techniques described by~\citet{lon10}. Although we
normalized our $s$-wave contribution to the experimental thermal value of $\sigma_\mathrm{th}$ = (0.67$\pm$0.07) mb, it deviates from the old value of (0.538$\pm$0.065) mb
by about 20\%, and thus we estimated conservatively the uncertainties in the $s$-wave contribution to be 20\%. The uncertainties in the $p$-wave contribution is estimated to be
50\% as discussed in the previous DC section. The uncertainties in the resonance contributions (including the subthreshold state) are determined by the parameter uncertainties as
listed in Table~\ref{tab1_par}. The uncertainties of the present results are shown as the colored error bands in Figs.~3--5.

\subsection{Total reaction rate}
\label{sec2.4}
\begin{figure}[t]
\begin{center}
\includegraphics[width=0.6\textwidth]{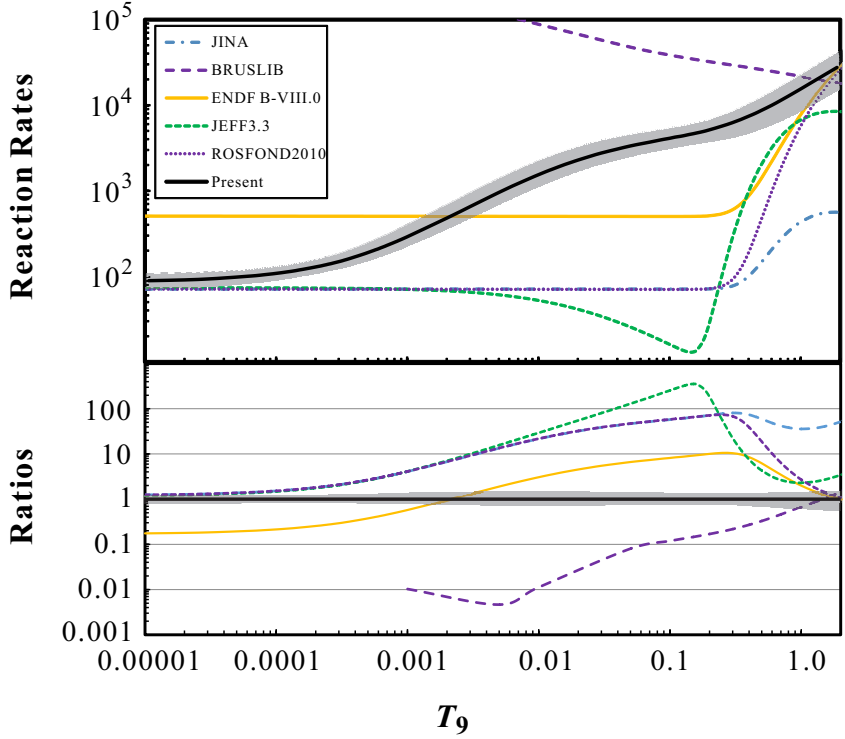}
\caption{\label{fig5} Thermonuclear $^{17}$O($n$,$\gamma$)$^{18}$O rates (in units of cm$^3$s$^{-1}$mol$^{-1}$) as a function of $T_9$: the upper panel for reaction rates, and the
lower panel for ratios between the present results and others. The uncertainties of the present rates indicated as the error band are calculated by a Monte-Carlo method.}
\end{center}
\end{figure}

Based on the presently calculated and evaluated cross sections shown in Fig.~\ref{fig4}, we have numerically calculated the thermonuclear $^{17}$O($n$,$\gamma$)$^{18}$O rates
using the approach discussed above. The reaction rates and the associated ratios between the present results and others are shown in Fig.~\ref{fig5}, respectively. If the misquoted
thermal value in ENDF/B-VIII is corrected, the corresponding rate should be almost the same as that of ROSFOND2010. Therefore, our new rate is remarkably (up to 2 orders of
magnitude) larger than those calculated by using the cross-section data evaluated in the current ENDF/B, JEFF and ROSFOND libraries, in the temperature range up to $T_9$ = 2.
However, we think that the BRUSLIB overestimated this rate significantly because of the present light target system of $^{17}$O. The JINA \emph{bb92} rate, which is the exact one
derived by~\citet{koe91}, is also small, because the DC $p$-wave capture and subthreshold state contributions were not taken into account in their calculations. In the temperature
region of $T_9$ = 0.1 $\sim$ 2, the present rate is about 35$\sim$80 times larger than the \emph{bb92} rate, which is frequently utilized in the nucleosynthesis network calculations.
The numerical values of the present reaction rate and the associated lower and upper limits are listed in Table~\ref{tab2_rate}, where other rates are also listed for comparison.

For completeness, the $^{17}$O($n$,$\gamma$)$^{18}$O rate in the high temperature region of $T_9$ = 2 $\sim$ 10 has been treated below. As shown in Fig.~\ref{fig4}, the present
cross sections are about 1.45 times smaller than those of ENDF/B-VIII in the energy region of 0.5--1 MeV. Therefore we adopt the ENDF/B-VIII data in the energy region of 1.0--20 MeV
by dividing a normalization factor of 1.45. The $^{17}$O($n$,$\gamma$)$^{18}$O rate is then calculated numerically based on the presently calculated cross section data ($\leq$1 MeV)
and the normalized ENDF/B-VIII data (1--20 MeV), by using the Monte-Carlo approach. The high temperature rates are also listed in Table~\ref{tab2_rate}.
The present $^{17}$O($n$,$\gamma$)$^{18}$O (Mean) rate can be parameterized by the standard format of~\citet{rau00},
\begin{eqnarray}
N_A\langle\sigma v\rangle_{(n,\gamma)} &=& \mathrm{exp}(11.1633-\frac{0.0400988}{T_9}+\frac{4.56334}{T_9^{1/3}}-6.40963T_9^{1/3}+0.317629T_9-0.0163527T_9^{5/3}+4.25928 \ln{T_9}) \nonumber \\
                          &+& \mathrm{exp}(55.7181-\frac{0.184283}{T_9}+\frac{17.6063}{T_9^{1/3}}-76.4751T_9^{1/3}+10.0633T_9-1.1705T_9^{5/3}+21.7074\ln{T_9}) \, ,
\label{eq8}
\end{eqnarray}
with a fitting error of less than 0.2\% over the temperature region of $T_9$ = 0.01 $\sim$ 10.

\begin{table*}
\footnotesize
\caption{\label{tab2_rate} Presently derived thermonuclear $^{17}$O($n$,$\gamma$)$^{18}$O rate, together with the lower and upper limits (in units of cm$^3$s$^{-1}$mol$^{-1}$). The
ENDF/B-VIII, JEFF3.3 and ROSFOND2010 rates are calculated based on the cross sections evaluated in the respective libraries, and the rates from JINA REACLIB and BURSLIB are
also listed for comparison.}
\begin{tabular}{ccccccccc}
\hline
$T_9$  & Present(Mean)  & Lower limit  & Upper limit & ENDF/B-VIII &  JEFF3.3  & ROSFOND2010 & JINA REACLIB & BRUSLIB \\
\hline
0.01 & 1.56E+03 & 1.14E+03 & 2.18E+03 & 5.03E+02 & 5.24E+01 & 7.06E+01 & 7.12E+01 & 8.74E+04 \\
0.02 & 2.31E+03 & 1.70E+03 & 3.18E+03 & 5.02E+02 & 4.13E+01 & 7.06E+01 & 7.12E+01 & 6.64E+04 \\
0.03 & 2.77E+03 & 2.06E+03 & 3.77E+03 & 5.02E+02 & 3.43E+01 & 7.05E+01 & 7.12E+01 & 5.72E+04 \\
0.04 & 3.08E+03 & 2.32E+03 & 4.15E+03 & 5.02E+02 & 2.94E+01 & 7.05E+01 & 7.12E+01 & 5.17E+04 \\
0.05 & 3.31E+03 & 2.51E+03 & 4.42E+03 & 5.02E+02 & 2.57E+01 & 7.05E+01 & 7.12E+01 & 4.80E+04 \\
0.06 & 3.50E+03 & 2.66E+03 & 4.65E+03 & 5.02E+02 & 2.29E+01 & 7.05E+01 & 7.12E+01 & 4.52E+04 \\
0.07 & 3.66E+03 & 2.79E+03 & 4.83E+03 & 5.02E+02 & 2.07E+01 & 7.05E+01 & 7.12E+01 & 4.31E+04 \\
0.08 & 3.80E+03 & 2.89E+03 & 4.98E+03 & 5.02E+02 & 1.88E+01 & 7.05E+01 & 7.12E+01 & 4.14E+04 \\
0.09 & 3.92E+03 & 2.98E+03 & 5.11E+03 & 5.02E+02 & 1.73E+01 & 7.05E+01 & 7.12E+01 & 4.00E+04 \\
0.10 & 4.04E+03 & 3.06E+03 & 5.23E+03 & 5.02E+02 & 1.61E+01 & 7.05E+01 & 7.12E+01 & 3.88E+04 \\
0.11 & 4.15E+03 & 3.13E+03 & 5.37E+03 & 5.02E+02 & 1.50E+01 & 7.05E+01 & 7.12E+01 & 3.78E+04 \\
0.12 & 4.26E+03 & 3.19E+03 & 5.49E+03 & 5.02E+02 & 1.42E+01 & 7.05E+01 & 7.12E+01 & 3.69E+04 \\
0.13 & 4.36E+03 & 3.25E+03 & 5.63E+03 & 5.02E+02 & 1.35E+01 & 7.05E+01 & 7.12E+01 & 3.61E+04 \\
0.14 & 4.46E+03 & 3.31E+03 & 5.76E+03 & 5.02E+02 & 1.31E+01 & 7.05E+01 & 7.12E+01 & 3.54E+04 \\
0.15 & 4.56E+03 & 3.36E+03 & 5.89E+03 & 5.03E+02 & 1.31E+01 & 7.06E+01 & 7.12E+01 & 3.47E+04 \\
0.16 & 4.65E+03 & 3.41E+03 & 6.02E+03 & 5.03E+02 & 1.36E+01 & 7.07E+01 & 7.12E+01 & 3.42E+04 \\
0.18 & 4.85E+03 & 3.51E+03 & 6.29E+03 & 5.06E+02 & 1.76E+01 & 7.11E+01 & 7.13E+01 & 3.31E+04 \\
0.20 & 5.04E+03 & 3.59E+03 & 6.58E+03 & 5.10E+02 & 2.76E+01 & 7.18E+01 & 7.15E+01 & 3.23E+04 \\
0.25 & 5.53E+03 & 3.81E+03 & 7.32E+03 & 5.37E+02 & 1.01E+02 & 7.68E+01 & 7.28E+01 & 3.06E+04 \\
0.30 & 6.05E+03 & 4.03E+03 & 8.09E+03 & 5.98E+02 & 2.75E+02 & 9.13E+01 & 7.71E+01 & 2.93E+04 \\
0.35 & 6.59E+03 & 4.27E+03 & 8.90E+03 & 7.05E+02 & 5.70E+02 & 1.25E+02 & 8.59E+01 & 2.82E+04 \\
0.40 & 7.14E+03 & 4.51E+03 & 9.76E+03 & 8.69E+02 & 9.74E+02 & 1.89E+02 & 1.00E+02 & 2.74E+04 \\
0.45 & 7.71E+03 & 4.78E+03 & 1.07E+04 & 1.10E+03 & 1.46E+03 & 2.97E+02 & 1.21E+02 & 2.66E+04 \\
0.50 & 8.30E+03 & 5.06E+03 & 1.16E+04 & 1.39E+03 & 2.01E+03 & 4.60E+02 & 1.46E+02 & 2.60E+04 \\
0.60 & 9.55E+03 & 5.64E+03 & 1.35E+04 & 2.19E+03 & 3.16E+03 & 9.76E+02 & 2.05E+02 & 2.49E+04 \\
0.70 & 1.08E+04 & 6.26E+03 & 1.54E+04 & 3.25E+03 & 4.26E+03 & 1.77E+03 & 2.70E+02 & 2.40E+04 \\
0.80 & 1.21E+04 & 6.89E+03 & 1.73E+04 & 4.56E+03 & 5.24E+03 & 2.84E+03 & 3.31E+02 & 2.32E+04 \\
0.90 & 1.34E+04 & 7.52E+03 & 1.93E+04 & 6.07E+03 & 6.05E+03 & 4.16E+03 & 3.86E+02 & 2.25E+04 \\
1.00 & 1.47E+04 & 8.15E+03 & 2.12E+04 & 7.76E+03 & 6.72E+03 & 5.70E+03 & 4.32E+02 & 2.19E+04 \\
1.25 & 1.80E+04 & 9.80E+03 & 2.61E+04 & 1.26E+04 & 7.80E+03 & 1.03E+04 & 5.12E+02 & 2.06E+04 \\
1.50 & 2.12E+04 & 1.14E+04 & 3.10E+04 & 1.79E+04 & 8.30E+03 & 1.55E+04 & 5.50E+02 & 1.96E+04 \\
1.75 & 2.45E+04 & 1.31E+04 & 3.58E+04 & 2.34E+04 & 8.46E+03 & 2.10E+04 & 5.62E+02 & 1.89E+04 \\
2.00 & 2.77E+04 & 1.48E+04 & 4.05E+04 & 2.89E+04 & 8.42E+03 & 2.67E+04 & 5.58E+02 & 1.82E+04 \\
2.50 & 3.40E+04 & 1.85E+04 & 4.94E+04 & 3.99E+04 & 8.08E+03 & 3.79E+04 & 5.27E+02 & 1.74E+04 \\
3.00 & 4.02E+04 & 2.27E+04 & 5.76E+04 & 5.05E+04 & 7.64E+03 & 4.87E+04 & 4.86E+02 & 1.70E+04 \\
4.00 & 5.23E+04 & 3.25E+04 & 7.21E+04 & 7.06E+04 & 6.81E+03 & 6.91E+04 & 4.09E+02 & 1.73E+04 \\
5.00 & 6.41E+04 & 4.34E+04 & 8.46E+04 & 8.94E+04 & 6.20E+03 & 8.83E+04 & 3.48E+02 & 1.93E+04 \\
6.00 & 7.58E+04 & 5.46E+04 & 9.64E+04 & 1.07E+05 & 5.79E+03 & 1.06E+05 & 3.01E+02 & 2.32E+04 \\
7.00 & 8.70E+04 & 6.54E+04 & 1.08E+05 & 1.25E+05 & 5.55E+03 & 1.24E+05 & 2.66E+02 & 3.02E+04 \\
8.00 & 9.83E+04 & 7.59E+04 & 1.19E+05 & 1.41E+05 & 5.43E+03 & 1.41E+05 & 2.39E+02 & 4.24E+04 \\
9.00 & 1.09E+05 & 8.61E+04 & 1.31E+05 & 1.58E+05 & 5.43E+03 & 1.57E+05 & 2.17E+02 & 6.39E+04 \\
10.00 & 1.20E+05 & 9.59E+04 & 1.43E+05 & 1.73E+05 & 5.55E+03 & 1.73E+05 & 1.99E+02 & 1.03E+05 \\
\hline
\end{tabular}
\end{table*}

\section{Astrophysical implications}
\label{sec4}
The main origin of the trans-iron elements is believed to be either $s$-process or $r$-process which proceeds under the neutron-rich conditions. In this section, we examine the impacts of our new $^{17}$O($n$,$\gamma$)$^{18}$O rate on the both nucleosynthetic processes.
The stellar models and the calculated results are described in the following subsections in details. Thereinto, the previous $^{17}$O($n$,$\gamma$)$^{18}$O rate used in the models refers to the
\emph{bb92} rate compiled in the JINA REACLIB.

\subsection{s-process}
The astrophysical site for the production of the $s$-process elements whose masses are lighter than $A<$ 90 is most likely the massive star, and this nucleosynthetic process is called weak $s$-process. On the other hand, those of heavier mass $A>$ 90 is believed to be produced in the light-to-intermediate mass AGB stars, which is called the main $s$-process.~\citep{koeppeler11}.

\subsubsection{Weak $s$-process in massive stars}

\begin{figure}[t]
\begin{center}
\includegraphics[width=12cm]{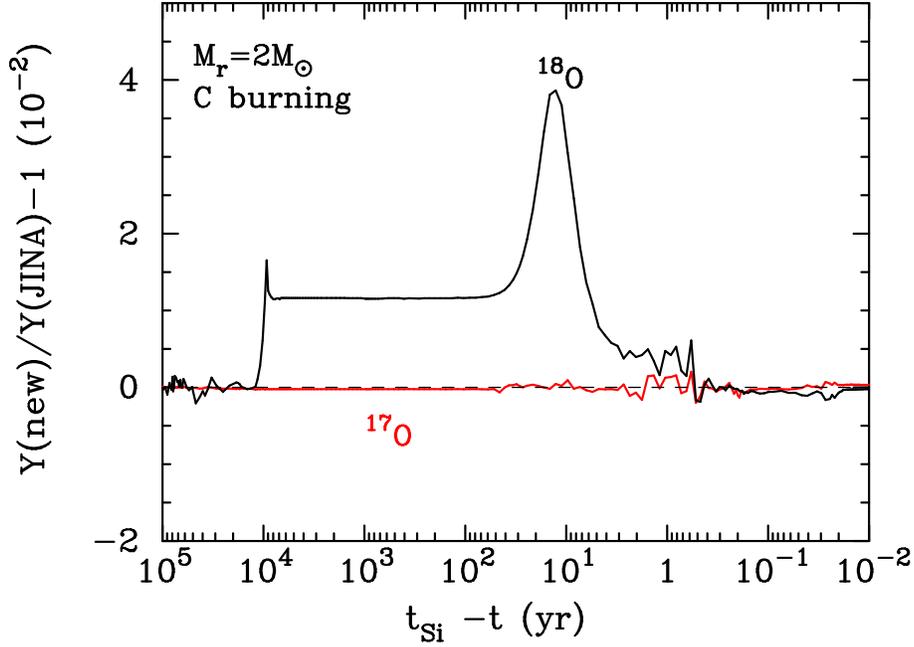}
\caption{\label{fig6} Deviations from unity of ratios of mole fractions of $^{17,18}$O in the cases of new and JINA rates as a function of the time to the start of core Si burning, i.e.,
$t_{\rm Si} -t$ during the shell C burning in a star with initial mass $M=25 M_\sun$ at $M_r$ = 2 $M_\odot$. See text for details.}
\end{center}
\end{figure}

In this subsection we study the impact of the present $^{17}$O($n$,$\gamma$)$^{18}$O rate on the evolution of light element abundances in massive stars. We adopt a
multi-zone nucleosynthesis calculation~\citep{HeM20} for a model star with $M$ = 25 $M_\sun$ from the Modules for Experiments in Stellar Astrophysics
(MESA)~\citep{pax11,pax15,pax18,pax19}. There is no discernable difference in nuclear abundances when a result for the new rate is compared with that of the JINA rate for both core He and shell C burning stages (see Figures 6 and 7 in~\citet{HeM20}). However, the change in the reaction rate leads to a small change of $^{18}$O abundance which is maximized
$\sim 10$ yr before the onset of Si burning. Figure~\ref{fig6} shows ratios of mole fractions $Y_i$ of $i$ = $^{17,18}$O at the Lagrangian mass coordinate $M_r =2 M_\sun$, for
example, as a function of the time to the start of core Si burning, i.e., $\Delta t \equiv t_{\rm Si} -t$. When the new rate is utilized, the $^{18}$O abundance increases by up to
$\sim 4$\% at $\Delta t \sim 10$ yr. At $\Delta t \sim 10^4$ yr in the last stage of He burning, a neutron exposure from $^{22}$Ne($\alpha$,$n$) enhances the $^{18}$O
abundances via the reaction $^{17}$O($n$,$\gamma$)$^{18}$O. In this epoch, the $^{18}$O production dominantly proceeds via $^{18}$F($\beta^+$)$^{18}$O,
$^{14}$C($\alpha$,$\gamma$)$^{18}$O, and $^{17}$O($n$,$\gamma$)$^{18}$O. Then, an increase of the last reaction rate results in a slight increase of the $^{18}$O abundance.
At $\Delta t \sim 10$ yr in the C burning stage, the neutron exposure is again gradually increasing due to $\alpha$ particles provided from the C+C fusion. Also in this epoch, the
$^{18}$O production occurs through $^{18}$F($\beta^+$)$^{18}$O, $^{14}$C($\alpha$,$\gamma$)$^{18}$O, and $^{17}$O($n$,$\gamma$)$^{18}$O, and an increase of $^{18}$O
abundance results again. At $\Delta t \lesssim 10$ yr, the rate of the reaction $^{17}$O($n$,$\gamma$)$^{18}$O is significantly smaller than the other 2 reactions for the $^{18}$O
production. Then, the effect of changing the $^{17}$O($n$,$\gamma$)$^{18}$O rate on the $^{18}$O abundance becomes smaller. We note that the effect of increasing $^{18}$O
abundance is cancelled at high temperatures since the other pathway of $^{17}$O($n$,$\alpha$)$^{14}$C($\alpha$,$\gamma$)$^{18}$O also contributes to the $^{18}$O yield when
enough $\alpha$ particles are supplied. Since the $^{17}$O abundance never becomes significantly large in non-rotating massive stars, the neutron capture reaction
$^{17}$O($n$,$\gamma$)$^{18}$O hardly affects the weak $s$-process.

\begin{figure}[t]
\begin{center}
\includegraphics[width=0.6\textwidth]{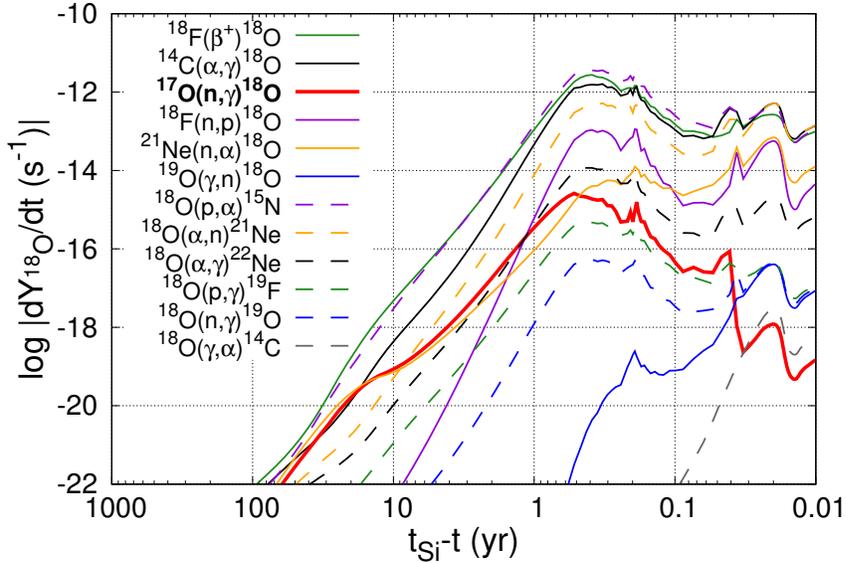}
\caption{\label{fig7} Reaction rates for production and destruction of $^{18}$O as a function of $t_{\rm Si} -t$ during the C burning in the same stellar model as in Fig.~\ref{fig6}.}
\end{center}
\end{figure}

Figure~\ref{fig7} shows rates for main production and destruction reactions of $^{18}$O as a function of $t_{\rm Si} -t$ during the C burning. The new rate of the
$^{17}$O($n$,$\gamma$)$^{18}$O reaction results in a $^{18}$O production via the reaction much more efficient than the old rate does. However, because the $\beta$-decay of $^{18}$F
and $^{14}$C($\alpha$,$\gamma$)$^{18}$O reaction are predominant in the $^{18}$O production, the sensitivity of $^{18}$O abundance to the $^{17}$O($n$,$\gamma$)$^{18}$O
reaction rate is still small. It can be seen that three reactions of $^{18}$O($p$,$\alpha$)$^{15}$N, $^{14}$C($\alpha$,$\gamma$)$^{18}$O and $^{18}$O($\alpha$,$n$)$^{21}$Ne
can most effectively influence the $^{18}$O abundance, except for the $^{18}$F $\beta$-decay.
At the same time, the impact of our new rate on heavier nuclear abundances involved in the present $s$-process network has been investigated, and such impacts are quite small,
only up to a level of ${\mathcal O}(10^{-3})$.

\subsubsection{Main $s$-process in AGB stars}
Utilizing a similar reaction network where the new and old rates of the reaction $^{17}$O($n$,$\gamma$)$^{18}$O are adopted, we investigate effects of new rate on elemental abundances during the main $s$-process in metal-poor stars in this subsection.
In our present model, we set up environmental conditions for nucleosynthesis in C- and $s$-enhanced metal-poor stars (MPSs), i.e., LP625-44 and CS31062-012 (equivalent to LP706-7). Observational data on abundances of $s$-elements are adopted from \citet{2006ApJ...650L.127A,2001ApJ...561..346A} for LP625-44
with [C/H] $=-0.43$\footnote{{\bf [A/B] $\equiv \log [(n_\mathrm{A}/n_\mathrm{B}) /(n_\mathrm{A}/n_\mathrm{B})_\sun]$, where $n_{i}$ is the number density of element $i$ and the mark $\sun$ indicates the solar value.}} and [Fe/H] $= -2.70$, and \citet{2008ApJ...678.1351A,2001ApJ...561..346A} for CS31062-012 with [C/H] $=-0.43$ and [Fe/H] = $-2.74$. Theoretical calculations of the main $s$-process are performed assuming repeated neutron capture episodes during interpulse phases of AGB stars~\citep[e.g.,][]{iwamoto04}. Solar abundances are taken from \citet{2009LanB...4B..712L}. For these two MPSs, a parametric study \citep{2001ApJ...561..346A} based upon a simple neutron injection model \citep{how86} demonstrates that the final yields only depend on the neutron exposure, i.e., $\tau_\mathrm{exp}=\int n_n v_T dt$, where $n_n$ is the neutron number density, and $v_T$ is the thermal neutron velocity at temperature $T$. Following this suggestion, we vary the neutron exposure assuming some specific setting in the present consistent network calculations.

The current $s$-process model assumes repeated episodes of flash-driven convective mixing~\citep{iwamoto04} followed by an interpulse phase that occurs under a radiative $s$-process condition with fixed temperature and density of $T_9=0.1$ and $\rho =500$~g cm$^{-3}$ \citep{1975ApJ...196..525I}.
  It is assumed that 10\% of the material that has experienced the $s$-process in $n$-th interpulse phase undergoes the $s$-process in the $(n+1)$-th interpulse phase, i.e., the overlap fraction $r=0.1$.
  Initial abundances in the $s$-process region are adopted so that they satisfy observed data, i.e., $^{12}$C/$^{13}$C isotopic ratios \citep{2001ApJ...561..346A,2002PASJ...54..933A} and abundance ratios of C, N, and O \citep{2001ApJ...561..346A,2002PASJ...54..933A,2008ApJ...678.1351A,2011ApJ...743..140B}: for LP625-44, $Y(^{12}\mathrm{C})=7.9 \times 10^{-3}$, $Y(^{13}\mathrm{C})=4.0 \times 10^{-4}$, $Y(^{14}\mathrm{N})=1.5 \times 10^{-4}$, and $Y(^{16}\mathrm{O})=7.9 \times 10^{-3}$; for CS31062-012, $Y(^{12}\mathrm{C})=7.8 \times 10^{-4}$, $Y(^{13}\mathrm{C})=7.8 \times 10^{-4}$, $Y(^{14}\mathrm{N})=1.5 \times 10^{-4}$, and $Y(^{16}\mathrm{O})=3.6 \times 10^{-3}$. Note that we need some adjustment for CS31062-012 to fit the heavy elemental abundance pattern. As a result, the total CNO abundance is rather small, and $^{13}$C abundance has been amplified by a factor of 15 to reproduce a large enough neutron exposure. In addition, the $^{22}$Ne mole fraction is set to the total mole fraction of CNO nuclei at the star formation~\citep{1975ApJ...196..525I}.
Since we fix the initial $^{13}$C abundance, the duration of interpulse phase is taken as a free parameter that determines the amount of neutron injection during the $s$-process. Effects of the $^{22}$Ne($\alpha$,$n$)$^{25}$Mg reaction during pulse phases are not considered. Although this reaction partially contributes to the $s$-process near the solar metallicity condition~\citep{2014JPhG...41e3101R} the $^{22}$Ne abundance is small in the low metallicity condition and the effect is expected to be negligibly small.

Initial nuclear abundances before the H-burning are assumed as follows. Below Fe-peak, abundances satisfy [X/Fe] = 0.3 \citep{1997ARA&A..35..503M} except for H and He that are taken at solar values because of their primordial origin. For Fe-peak elements, scaled solar abundances~\citep{2009LanB...4B..712L} of [X/Fe] = 0 are taken. For nuclei heavier than the Fe-peak elements, scaled $r$-elemental abundances are adopted, i.e., [r/Fe] = 0. Relative contributions from the $s$- and $r$-processes to solar abundances are adopted from~\citet{2014ApJ...787...10B}. The initial $s$-elemental abundances are set to zero since there is no enough time for evolution of low and intermediate mass stars that provide $s$-nuclei before the formation of such low-metallicity stars at very early cosmic time. Although enrichment of $r$-nuclei in the early epoch is uncertain~\citep{yam21}, results of $s$-process do not depend on the initial heavy element abundances~\citep{2001ApJ...561..346A}.

We have derived the best fit abundance patterns for the $s$-nuclei by a chi-squared analysis. The parameters are the interpulse duration $\Delta t$ and [s/C], which is the overall normalization factors for all $s$-nuclei applied to the calculated results of nucleosynthesis. In the $s$-process region, the hydrogen abundance is very small, and the use of the ratio [X/H] is in convenient. Instead, the $s$-process in AGB stars is associated with abundant C and is known as a partial C producer in Galactic chemical evolution~\citep{2000ApJ...541..660H}. Therefore, we take the ratio [X/C] of the calculated results and make comparisons with observed abundance ratios.

\begin{figure}
\begin{center}
\includegraphics[width=0.51\textwidth]{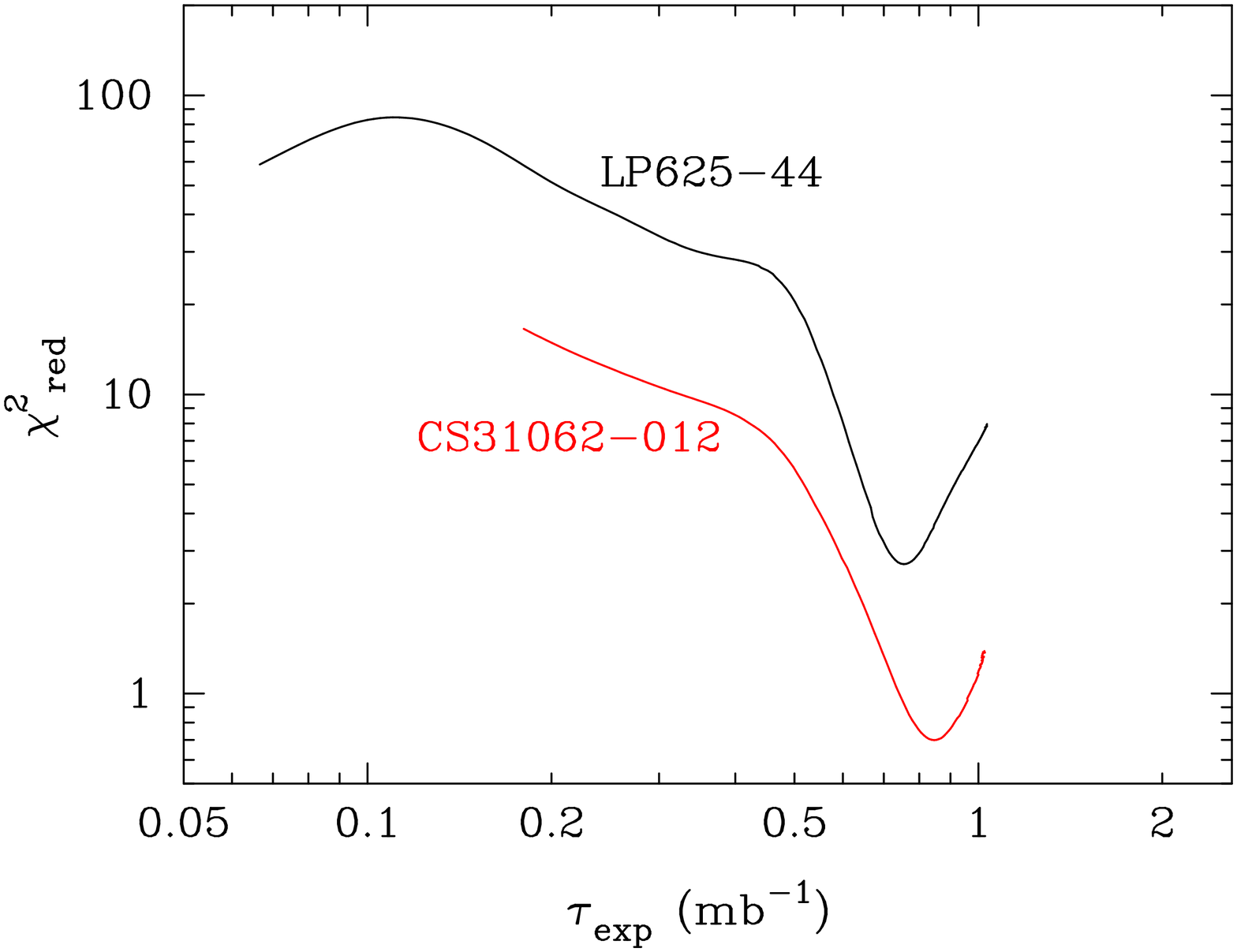}
\includegraphics[width=0.51\textwidth]{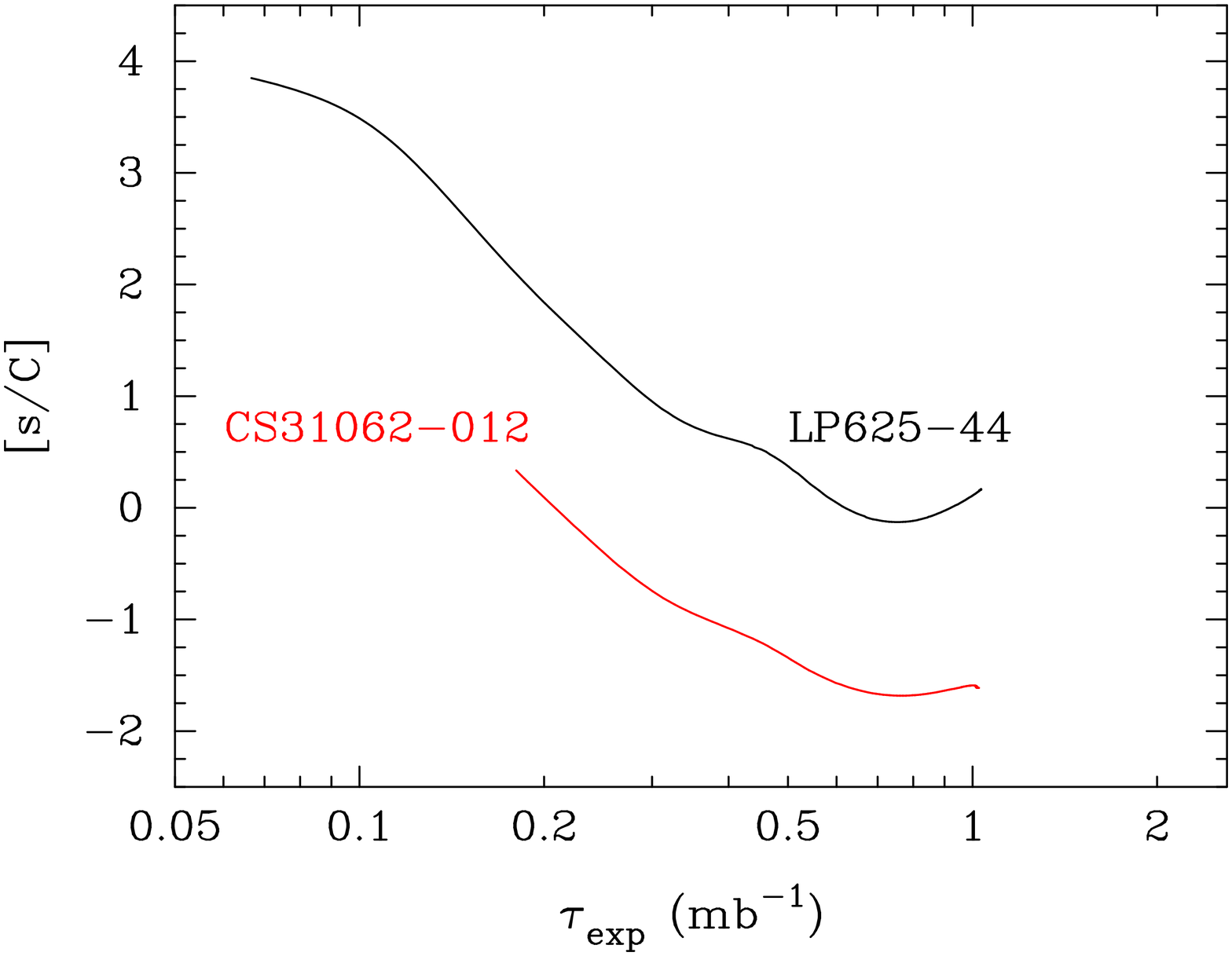}
\includegraphics[width=0.51\textwidth]{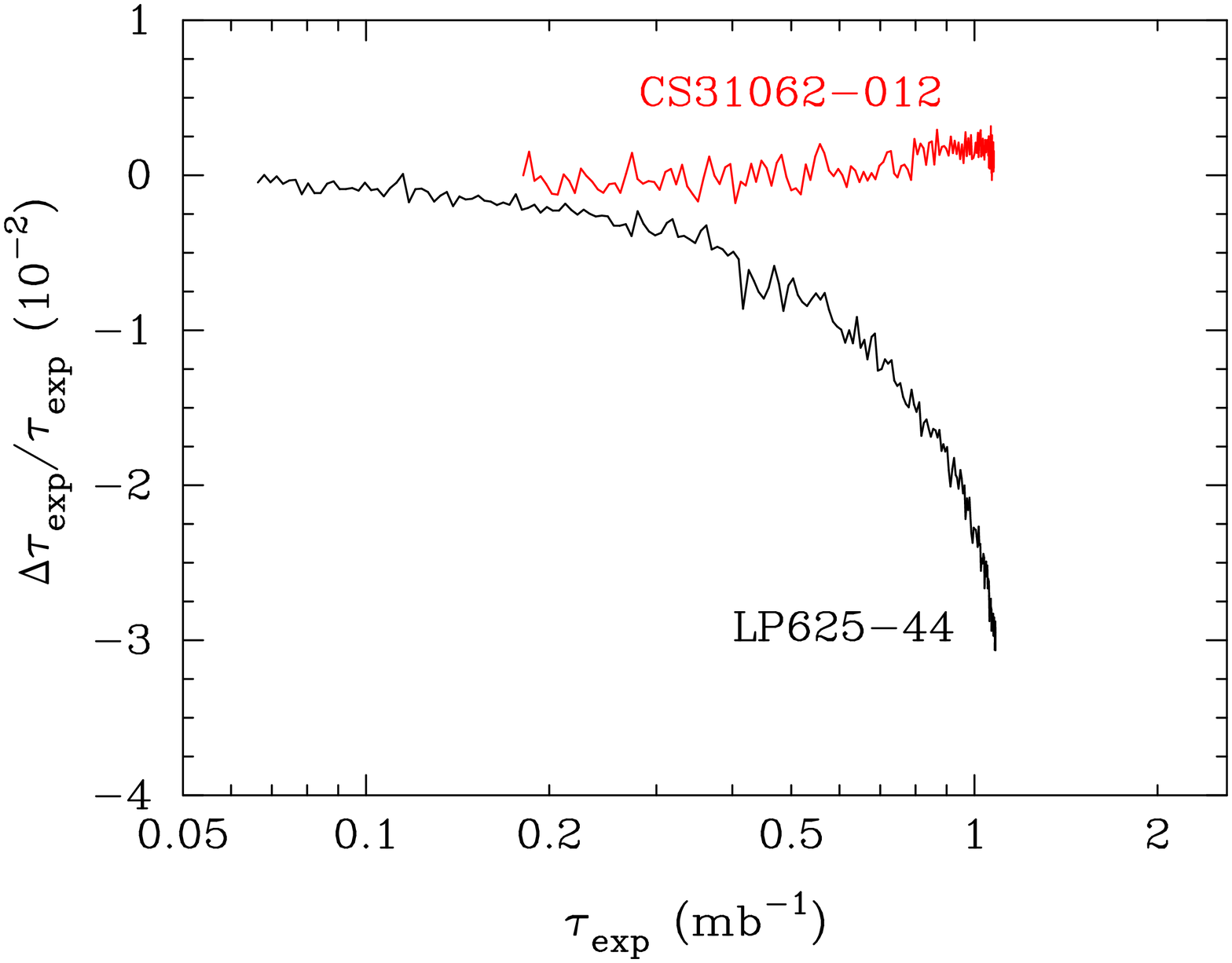}
\end{center}
\caption{Reduced $\chi^2$ versus the neutron exposure for the best fit cases (top panel) and fitted normalization factor versus the neutron exposure (middle panel). The normalization factor was added to logarithmic abundance ratio of [X/C] to fit the observed abundances. Also shown are fractional differences in the neutron exposure between cases with the new and old rates of $^{17}$O($n$,$\gamma$)$^{18}$O (bottom panel). Results for LP625-44 and CS31062-012 are shown.
\label{fig8}}
\end{figure}

Figure~\ref{fig8} shows results of fittings for the two MPSs. The upper panel shows the reduced $\chi^2$ at the best fit normalization factor as a function of the neutron exposure. The fitted normalization factor is shown in the middle panel. Both stars have local minima at $\tau_\mathrm{exp} \lesssim 1$ mb$^{-1}$ where the normalization factors are minimal. When the $\tau_\mathrm{exp}$ value of LP625-44 is smaller or larger, the required normalization factor increases, indicating that those cases do not correspond to the nucleosynthesis in LP625-44. Because of the minimum normalization factor close to no need for abundance shifts relative to C abundance, i.e., [s/C] $\sim 0$, it is apparent that the $\chi^2$ minimum is the best fit case in the adopted model. For CS31062-012, the large $|$[s/C]$|$ value indicates the diffuculty in reproduction of the $s$-process in that star with the one-zone model. The best cases correspond to the following values:
$\tau_\mathrm{exp} = 0.75$ mb$^{-1}$, [s/C]$ =-0.13$, and $\Delta t =16$ kyr for LP625-44;
$\tau_\mathrm{exp} =0.85$ mb$^{-1}$, [s/C]$ =-1.7$, and $\Delta t =7.6$ kyr for CS31062-012.


Figure~\ref{fig8} (botom panel) shows differences in $\tau_\mathrm{exp}$ caused by the use of the different reaction rates for $^{17}$O($n$,$\gamma$)$^{18}$O. Within the parameter region of the metal-poor stars searched in this study, ${\mathcal O}(1)$ \% of differences are found. This difference sensitively depends on the neutron exposure (horizontal axis) and initial nuclear abundances. Therefore, it is important to determine reaction rates that have been unmeasured in order to understand the main $s$-process under metal-poor conditions.

\begin{figure}
\begin{center}
\includegraphics[width=0.51\textwidth]{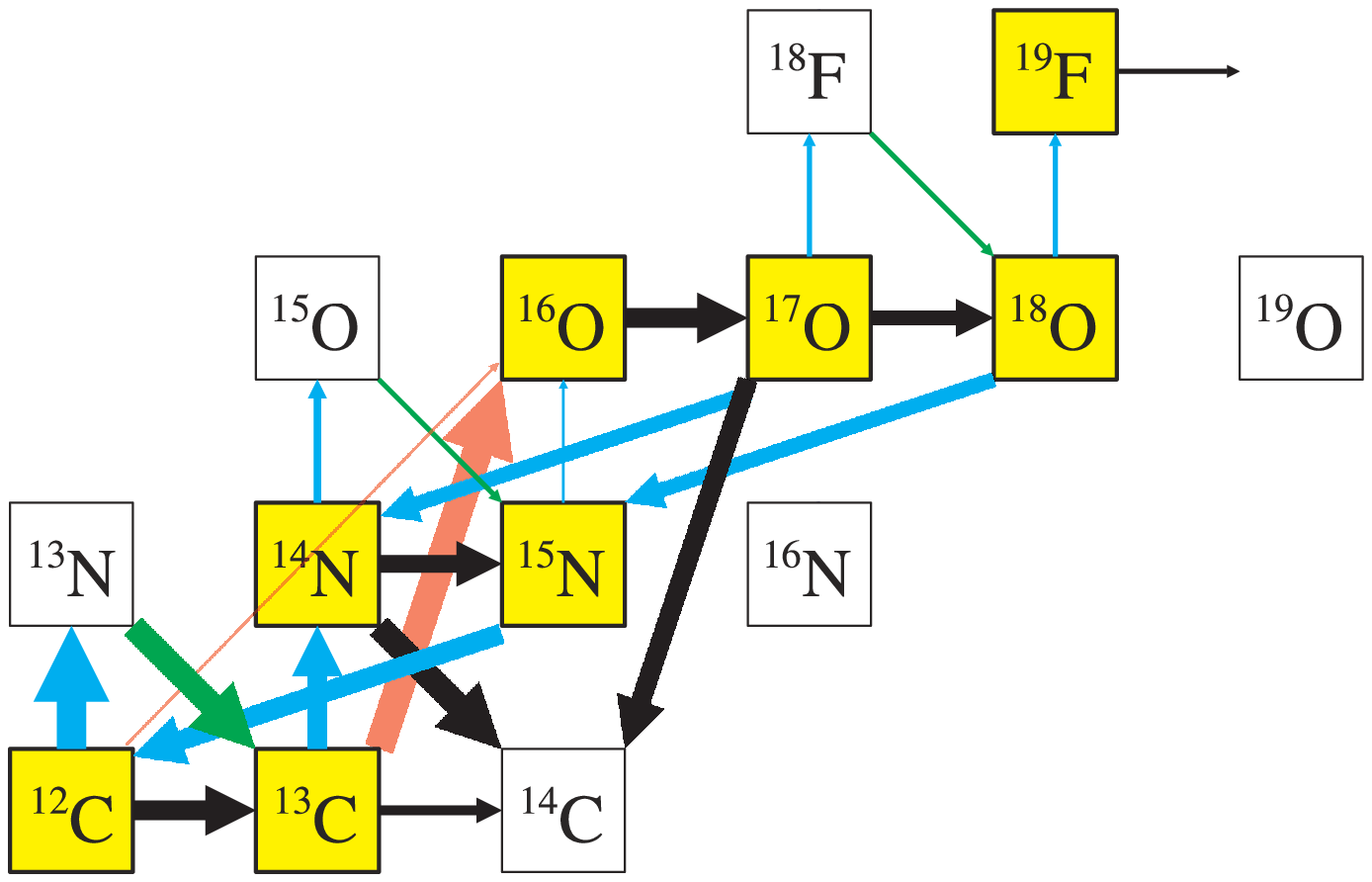}
\end{center}
\caption{Nuclear flows in the $s$-process via main reactions in the best case for LP625-44 with the new rate. Arrows correspond to reactions of ($n$,$\gamma$), ($n$,$p$), ($n$,$\alpha$), ($\alpha$,$n$), ($p$,$\alpha$), ($p$,$\gamma$) and $\beta$-decays. Each arrow is shown for the flow, i.e., $\Delta Y_{ij}$ for nuclei $i$ and reaction $j$ if $\Delta Y_{ij} > 10^{-8}$ with the thickness proportional to $\log (\Delta Y_{ij} / 10^{-8})$. Stable nuclei are shown by the yellow shaded boxes.
\label{fig9}}
\end{figure}

Figure~\ref{fig9} shows nuclear flows in the $s$-process via main reactions in the best case for LP625-44 with the new rate. The thickness of each arrow reflects the logarithmic amplitude of the flow, i.e., $\Delta Y_{ij}$ for nuclei $i$ and reaction $j$. The dominant neutron source is $^{13}$C($\alpha$,$n$)$^{16}$O ($\Delta Y =2.1 \times 10^{-3}$). The dominant neutron poison is $^{14}$N via $^{14}$N($n$,$p$)$^{14}$C ($\Delta Y =1.1 \times 10^{-3}$). There is a significant flow via $^{17}$O($n$,$\gamma$)$^{18}$O ($\Delta Y =1.1 \times 10^{-5}$). Because of protons produced abundantly via $^{14}$N($n$,$p$)$^{14}$C, $^{18}$O is mainly destroyed via $^{18}$O($p$,$\alpha$)$^{15}$N ($\Delta Y =1.1 \times 10^{-5}$). $^{15}$N nuclei tend to be destroyed via $^{15}$N($p$,$\alpha$)$^{12}$C ($\Delta Y =6.5 \times 10^{-5}$). In metal-poor stars the dominant neutron-poison is the light nuclei, not the minor heavier nuclei including Fe-peak elements. Thus, a large change in the $^{17}$O($n$,$\gamma$)$^{18}$O rate influences the neutron budget among the light nuclei, and the neutron exposure is noticeably revised as seen in Fig.~\ref{fig8} (botom panel).
We note that $^{19}$F is produced in the interpulse phase predominatly via $^{18}$O($p$,$\gamma$)$^{19}$F ($\Delta Y =6.4 \times 10^{-8}$), not via $^{18}$O($n$,$\gamma$)$^{19}$O ($\Delta Y =8.5 \times 10^{-10}$) followed by the $\beta^-$ decay.


\begin{figure}
\begin{center}
\includegraphics[width=0.6\textwidth]{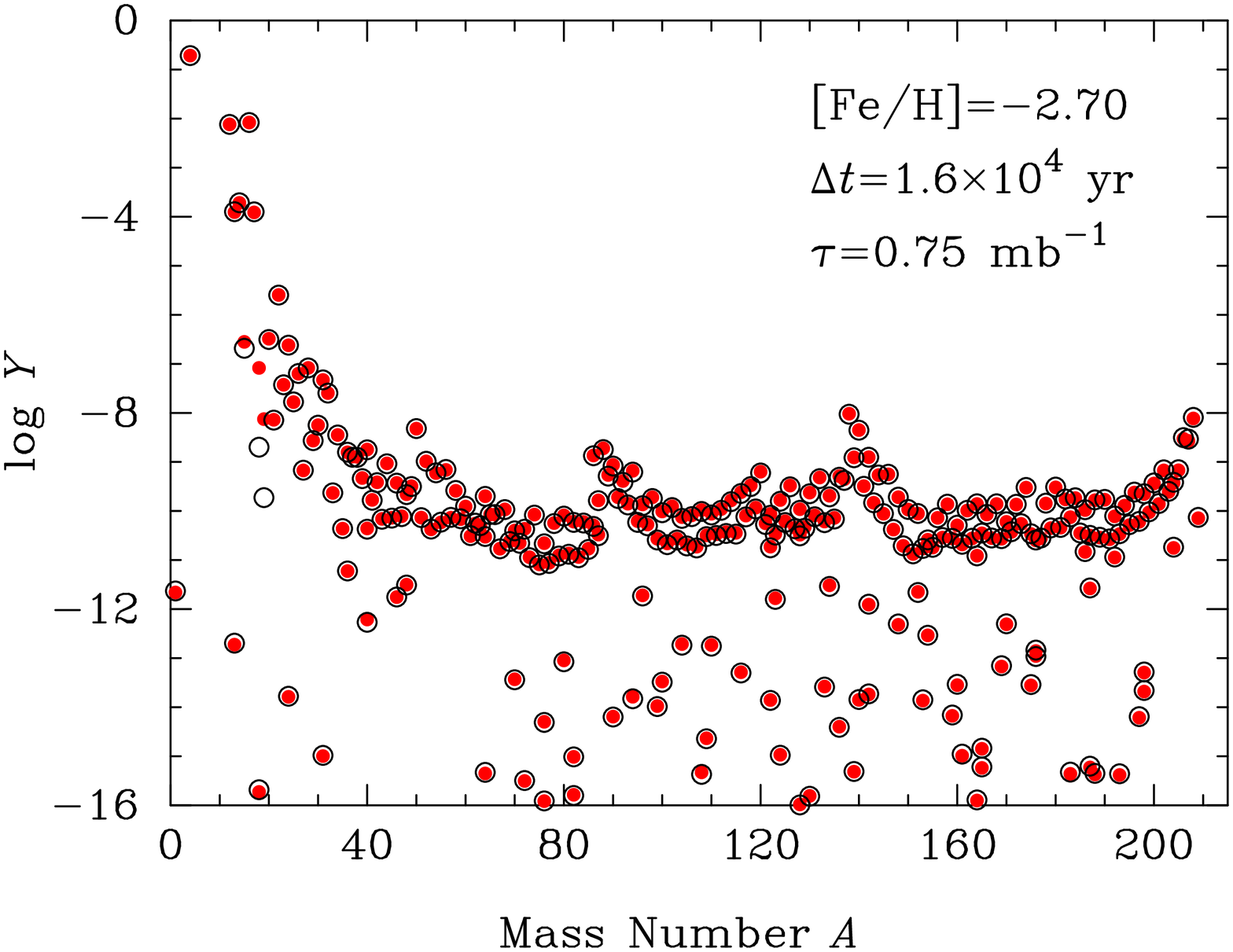}
\includegraphics[width=0.6\textwidth]{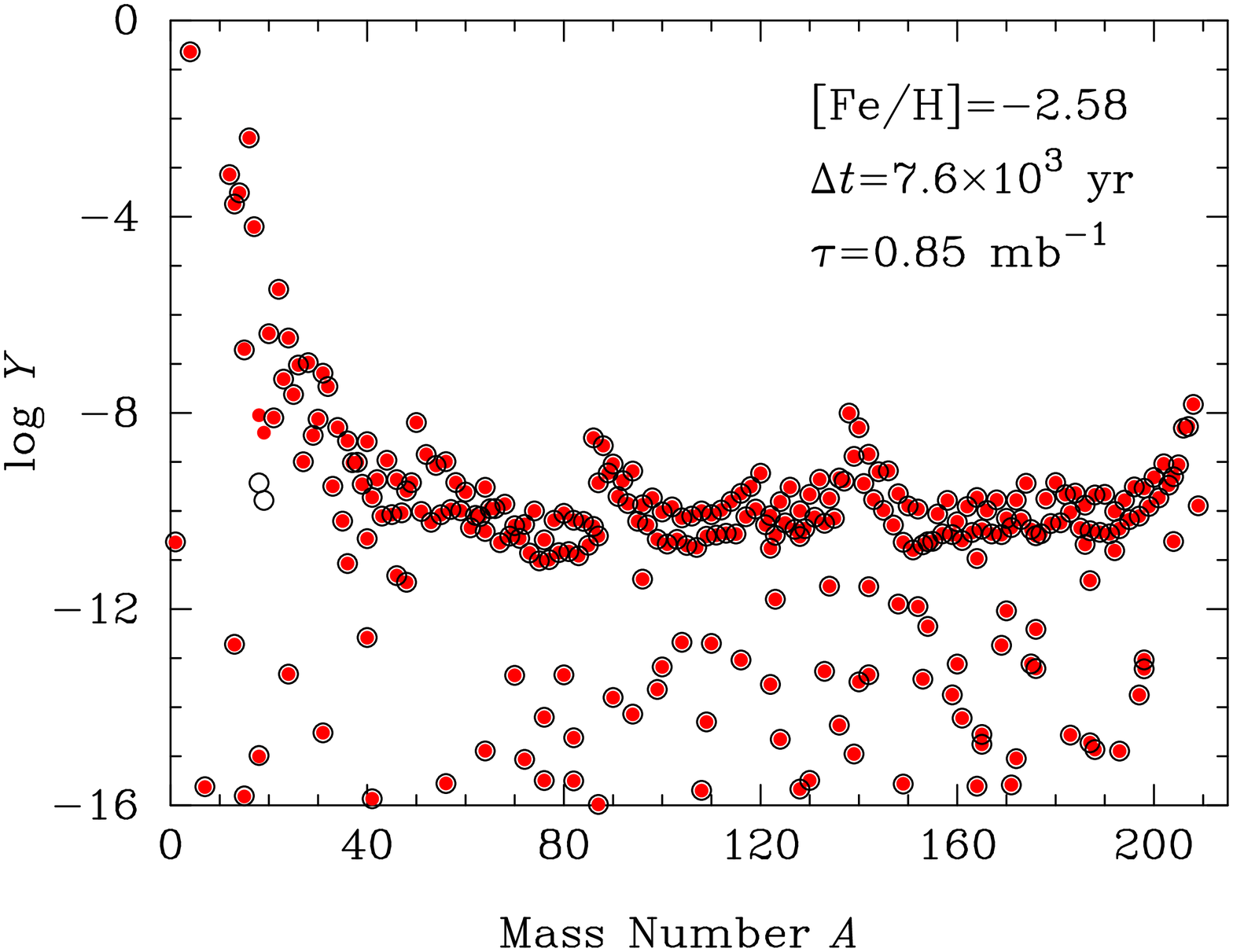}
\end{center}
\caption{Nuclear abundances after realization of converging abundances of $s$-only nuclei in the $s$-process for metallicities of LP625-44 (upper panel) and CS31062-012 (lower), respectively. Solid and open circles correspond to the new and old rates for the reaction $^{17}$O($n$,$\gamma$)$^{18}$O, respectively, in the best fit models.
\label{fig10}}
\end{figure}

\begin{figure}
\begin{center}
\includegraphics[width=0.6\textwidth]{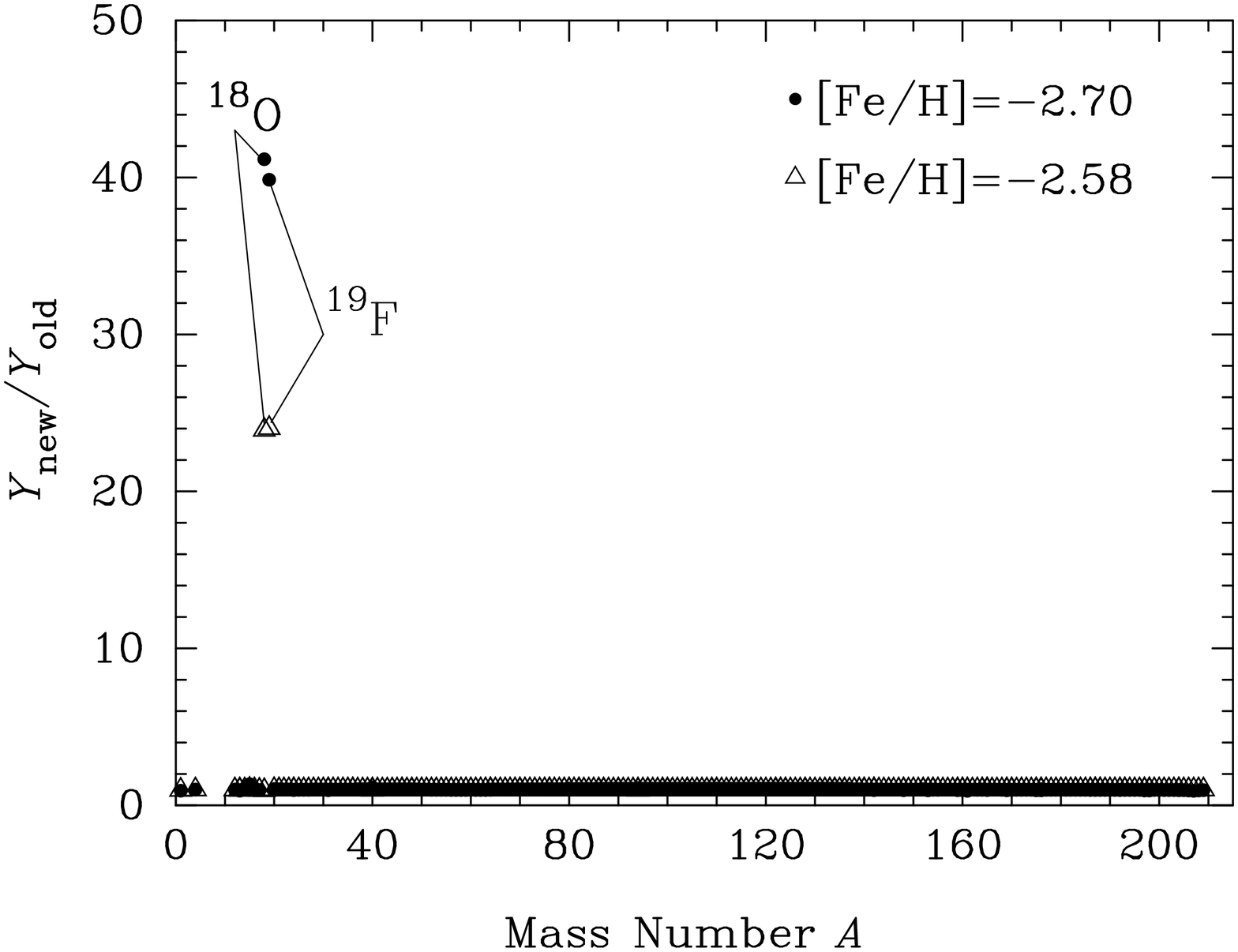}
\end{center}
\caption{Ratios of nuclear yields of the $s$-process with the new and old reaction rates of $^{17}$O($n$,$\gamma$)$^{18}$O, for metallicities of LP625-44 (solid circles) and CS31062-012 (open triangles), respectively.
\label{fig11}}
\end{figure}

Figure~\ref{fig10} shows final yields of the $s$-process in the best cases as a function of mass number. The final yields are identified as the yields after the first interpulse phase after which abundances of $s$-only nuclei reproduce those after the previous interpulse. The new rate of $^{17}$O($n$,$\gamma$)$^{18}$O corresponds to the solid circles, and the old rate corresponds to the open circles. Figure \ref{fig11} shows ratios for the final yields of the $s$-process as a function of mass number.
Noticeable changes are observed only in the $^{18}$O and $^{19}$F abundances. The newer rate leads to the dramatically enhanced abundances of $^{18}$O and $^{19}$F, by factors of 41 and 40, respectively, for LP625-44 and
by a factor of 24 for CS31062-012.


\begin{figure}
\begin{center}
\includegraphics[width=0.6\textwidth]{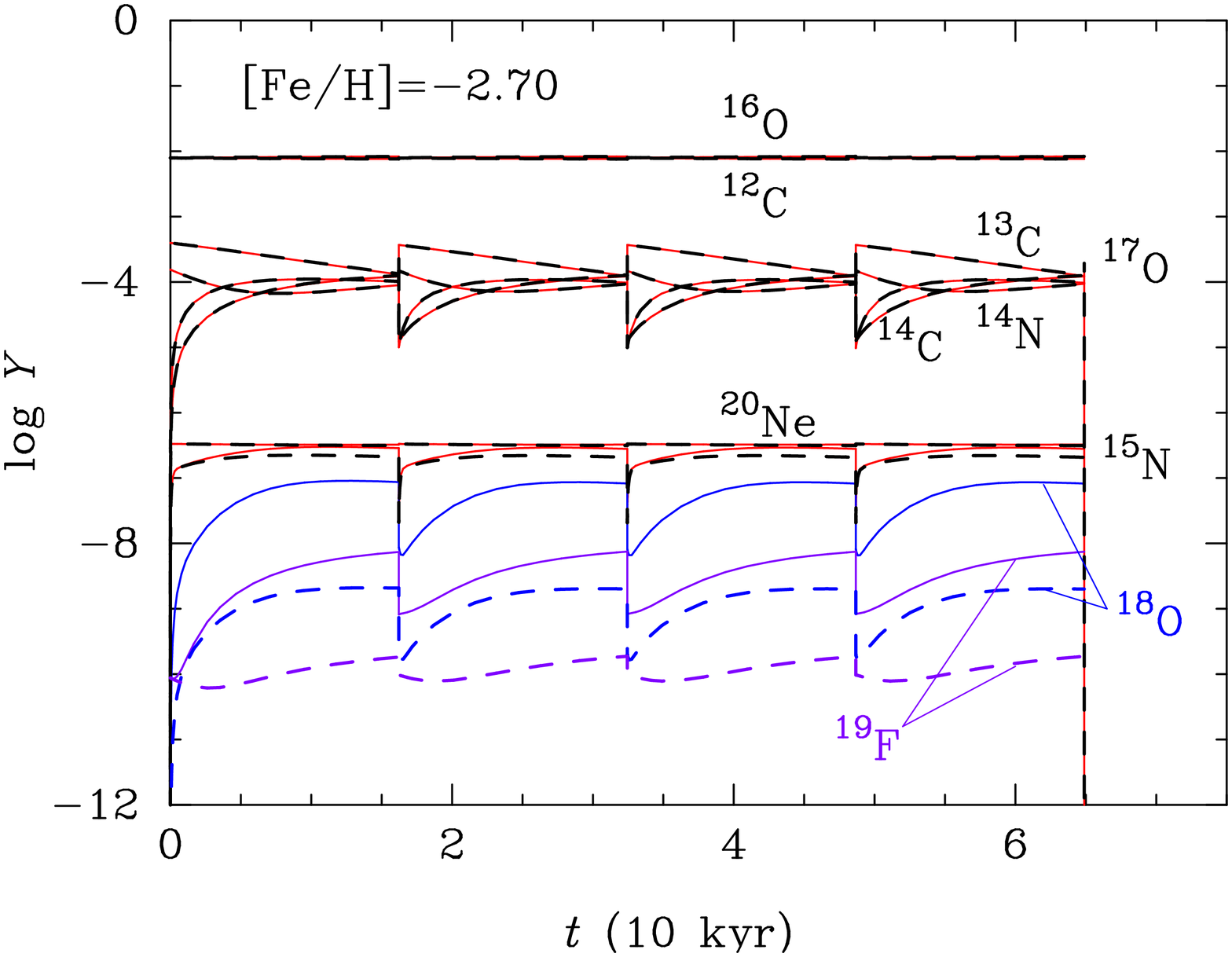}
\end{center}
\caption{Time evolution of nuclear abundances in the best fit model for LP625-44. Thin solid and thick dashed lines correspond to the new and old rates for the reaction $^{17}$O($n$,$\gamma$)$^{18}$O, respectively.
\label{fig12}}
\end{figure}

Figure \ref{fig12} shows time evolution of nuclear abundances in the best fit model for LP625-44 for the new (thin solid lines) and old (thick dashed lines) reaction rates. The convergence of $s$-process yields is obtained after four interpulse episodes. In the old rate case, production of $^{18}$O is very small, and the $^{19}$F abundance decreases after the mixing of material until the $^{18}$O abundance increases enough. However, in the new rate case, $^{18}$O abundance increases significantly via the larger $^{17}$O($n$,$\gamma$)$^{18}$O rate, and accordingly, the $^{19}$F abundance is more resistant to destruction due to the production via the $^{18}$O($p$,$\gamma$)$^{19}$F. These enhancements result in the differences as shown in Fig.~\ref{fig10}.


\begin{figure}
\begin{center}
\includegraphics[width=0.6\textwidth]{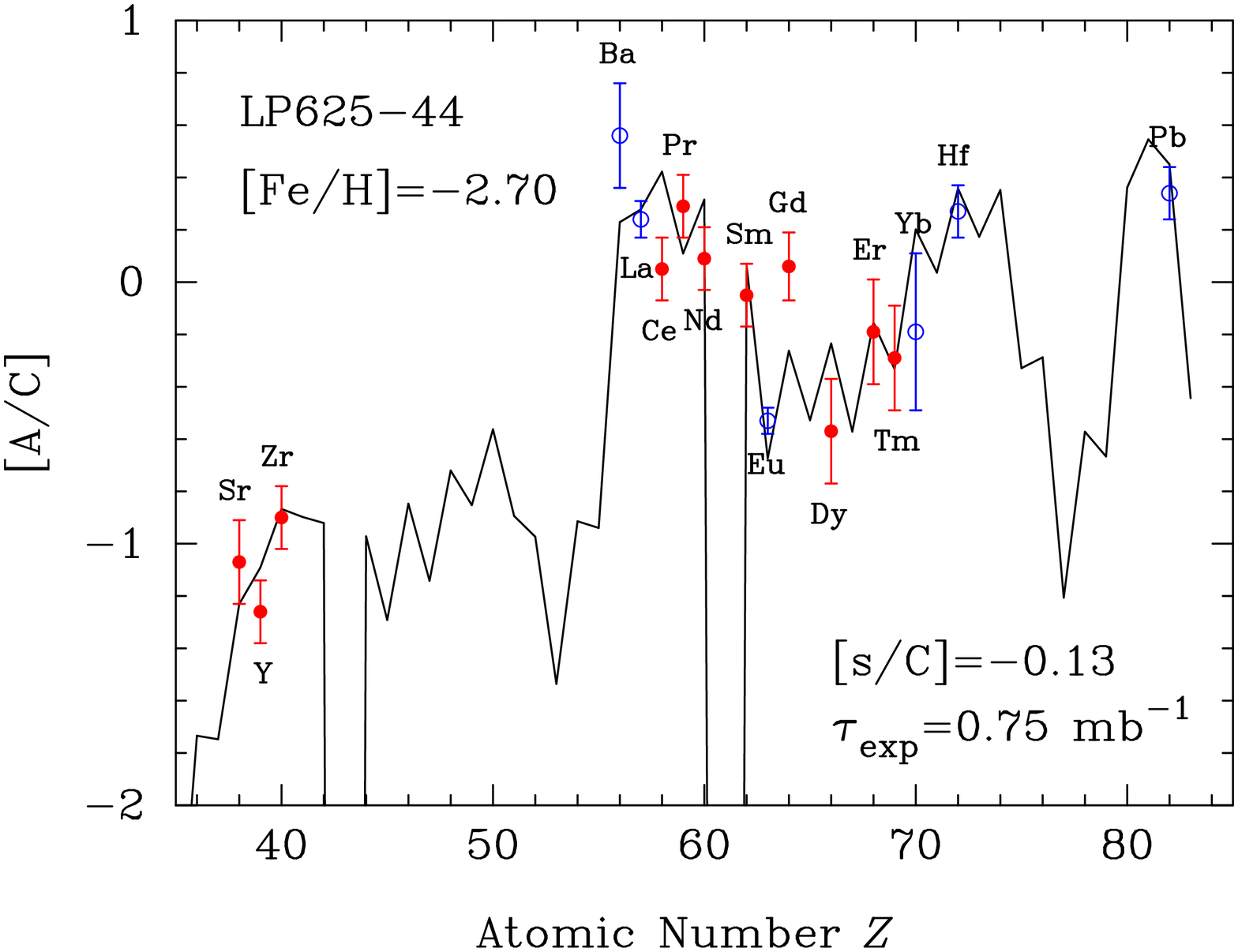}
\includegraphics[width=0.6\textwidth]{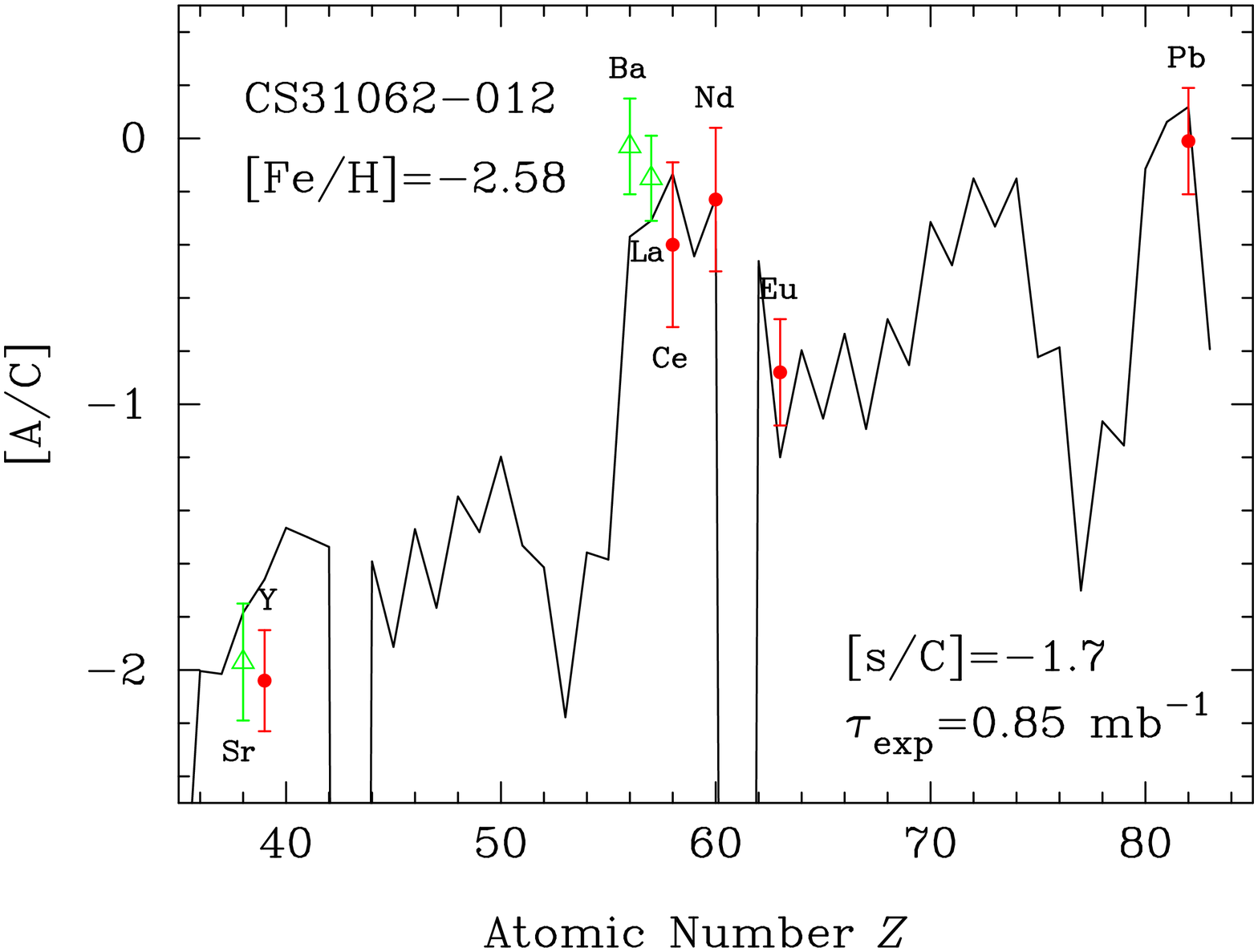}
\end{center}
\caption{Calculated best fit abundances of $s$-nuclei (lines) and observational data (points with error bars) as a function of mass number for LP625-44 (upper panel) and CS31062-012 (lower). Observational data are adopted from \citet{2001ApJ...561..346A} (solid circles), \citet{2008ApJ...678.1351A} (open circles), and
\citet{2006ApJ...650L.127A} (open triangles).
\label{fig13}}
\end{figure}

Figure~\ref{fig13} shows abundance ratios [A/C] of $s$-nuclei for calculated best fit abundances (lines) and observational data (points with error bars) for LP625-44 (upper panel) and CS31062-012 (lower panel).
The fittings are fairly good as indicated by small reduced $\chi^2$ values (Fig.~\ref{fig8}). The density, temperature, and initial nuclear abundances were fixed in the current calculation. Nevertheless, the best fit cases do not require any significant shifts of the calculated abundances of $s$-nuclei with respect to C abundance, i.e. [s/C]$\sim 0$. This is a piece of evidence that the environmental condition of the $s$-process in LP625-44 is well modeled without any normalization or artificial change in adopted physical quantities.


We note that in rotating stellar models effective diffusions produce abundant primary $^{14}$N nuclei that induce a burst of neutrons via the $^{14}$N($\alpha$,$\gamma$)$^{18}$F($\beta^+$)$^{18}$O($\alpha$,$\gamma$)$^{22}$Ne($\alpha$,$n$)$^{25}$Mg sequence~\citep{2018ApJS..237...13L}. Therefore, the larger rate of $^{17}$O($n$,$\gamma$)$^{18}$O may affect yields of $^{19}$F and $s$-process isotopes in rotating stars.

\subsection{$r$-process}
After the discovery of gravitational waves (GWs) from the binary neutron star merger (NSM) GW170817 and its associated kilonova GRB170817~\citep{abb17a,abb17b}, there is a growing consensus that the NSM could be a dominant astrophysical site for $r$-process
elements as well as CCSNe~\citep{shi16,kaj19}. In addition, collapsar also has been suggested to be another viable site for the $r$-process abundances in the Milky Way~\citep{sie19}.
NSM has an unavoidable time delay of cosmological time scale from the birth of progenitor stars until they merge due to very slow GW radiation of losing orbital energy. Collapsar and
CCSNe, on the other hand, quickly evolve in the stellar evolution time scale $\sim$10$^6$~y and contribute to enriching the interstellar materials. Taking account of all these candidate
sites, recent theoretical study of Galactic chemical evolution of $r$-process elements has proved that the collapsars and CCSNe could be the dominant contributor to the $r$-process abundances
of the Milky Way during entire history of Galactic evolution, while the NSM contribution has delayed and arrived recently~\citep{yam21}.

From the nuclear physics viewpoint, it is widely accepted concept of the $r$-process nucleosynthesis that the outflows from any explosive events contain only protons, neutrons and a small amount of alpha particles as
the initial composition. This does not apply to the r-process in dynamical ejecta from NSMs. Alpha particles and nucleons combine to form carbon, nitrogen, oxygen and heavier nuclei through the $\alpha$-process coupled with efficient neutron-capture flow followed by
$\beta$-decays. We therefore discuss the impact of the $^{17}$O($n$,$\gamma$)$^{18}$O new rate on the $r$-process nucleosynthesis in stellar evolution of massive stars and explosion based on codes in NucNet Tools~\citep{mey12}.

\subsubsection{$r$-process in collapsar}
Collapsar is a single massive star which literally culminates its evolution in collapsing into a black hole instead of neutron star in ordinary core-collapse
supernova, i.e. neutrino-driven wind or magneto-hydrodynamically driven jet supernova.
In view of the progress in understanding the
origin of $r$-process elements in the Milky Way, we first demonstrate the effect of our new reaction rate for $^{17}$O($n$,$\gamma$)$^{18}$O on the collapsar $r$-process in this subsection.

\begin{figure}[t]
\begin{center}
\includegraphics[width=0.6\textwidth]{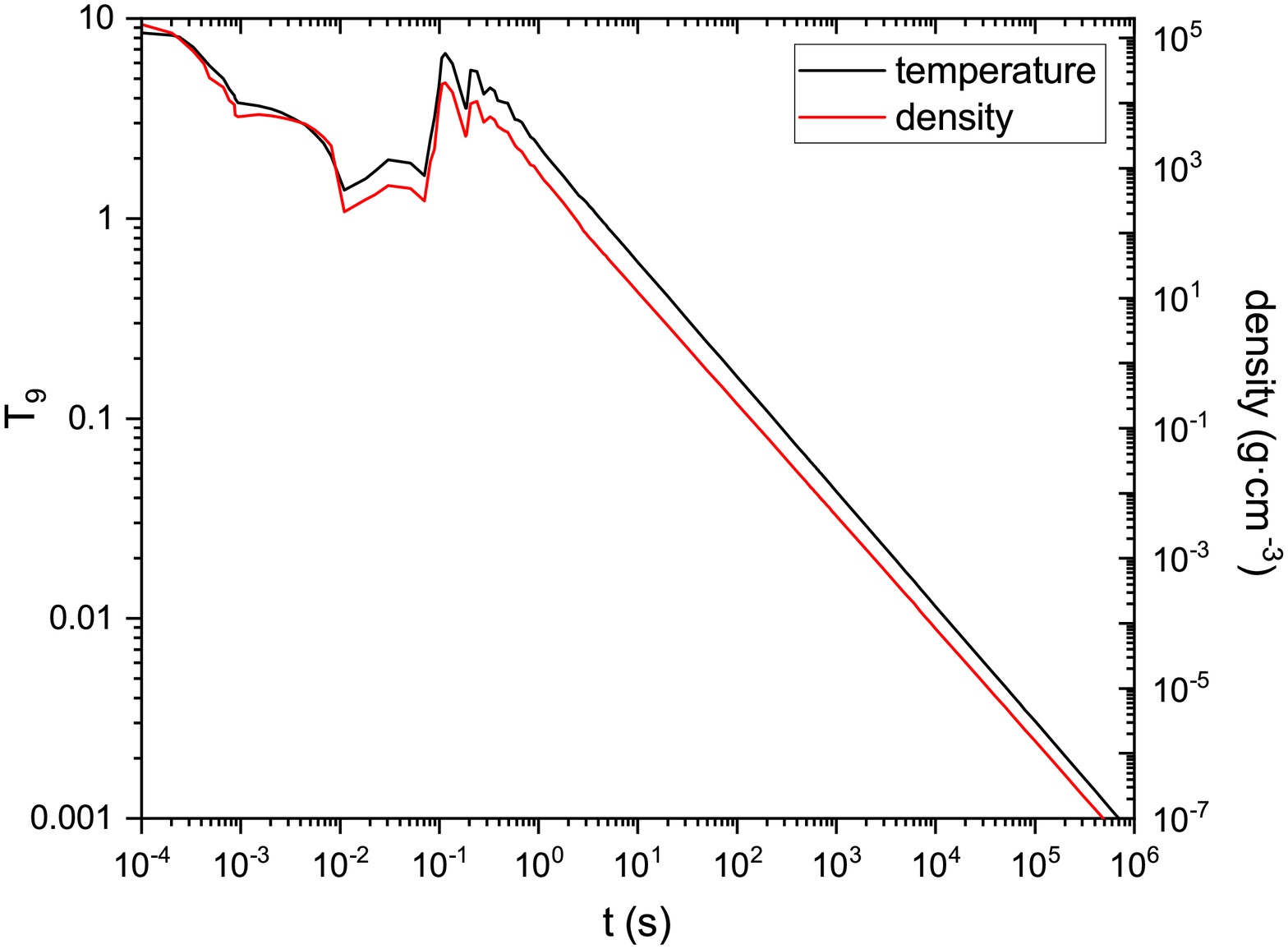}
\caption{\label{fig14} Temperature and density profiles used in the present $r$-process collapsar model.}
\end{center}
\end{figure}

The collapsar model which we use in our calculation is based on~\citet{har09}, whose progenitor mass, rotational velocity, and metallicity are 35 $M_\sun$, $v_\phi$ = 380 km/s, and
0.1~$Z_\sun$, respectively. We use the temperature and density profiles of one trajectory as shown in Fig.~\ref{fig14} from~\citet{nak15} for $r$-process nucleosynthesis. This trajectory
was used also in~\citet{fam20} for their theoretical studies of the effects of strong magnetic field on the $r$-process nucleosynthesis because it is the most effective and typical flow
among many trajectories to reach the fissile region.
In the following calculations, an initial electron mole fraction value of $Y_e$ = 0.05~\citep{nak15} has been utilized unless noted otherwise.

\begin{figure}[t]
\begin{center}
\includegraphics[width=0.6\textwidth]{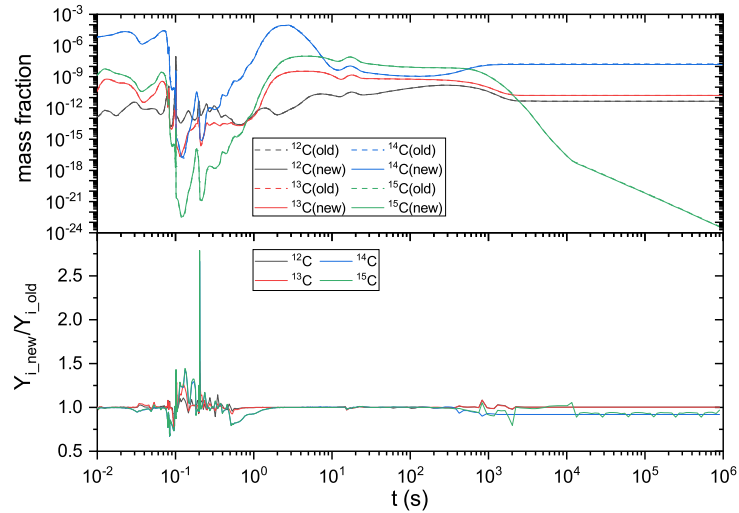}
\caption{\label{fig15} Time evolution of carbon isotopic abundances predicted in the $r$-process collapsar model ($Y_e$ = 0.05, without fission recycling effect). (Upper panel)
abundances calculated with the present (solid lines) and previous (dashed lines) $^{17}$O($n$,$\gamma$)$^{18}$O rates, respectively; (Lower panel) the corresponding abundance
ratios calculated using two rates.}
\end{center}
\end{figure}

\begin{figure}[t]
\begin{center}
\includegraphics[width=0.6\textwidth]{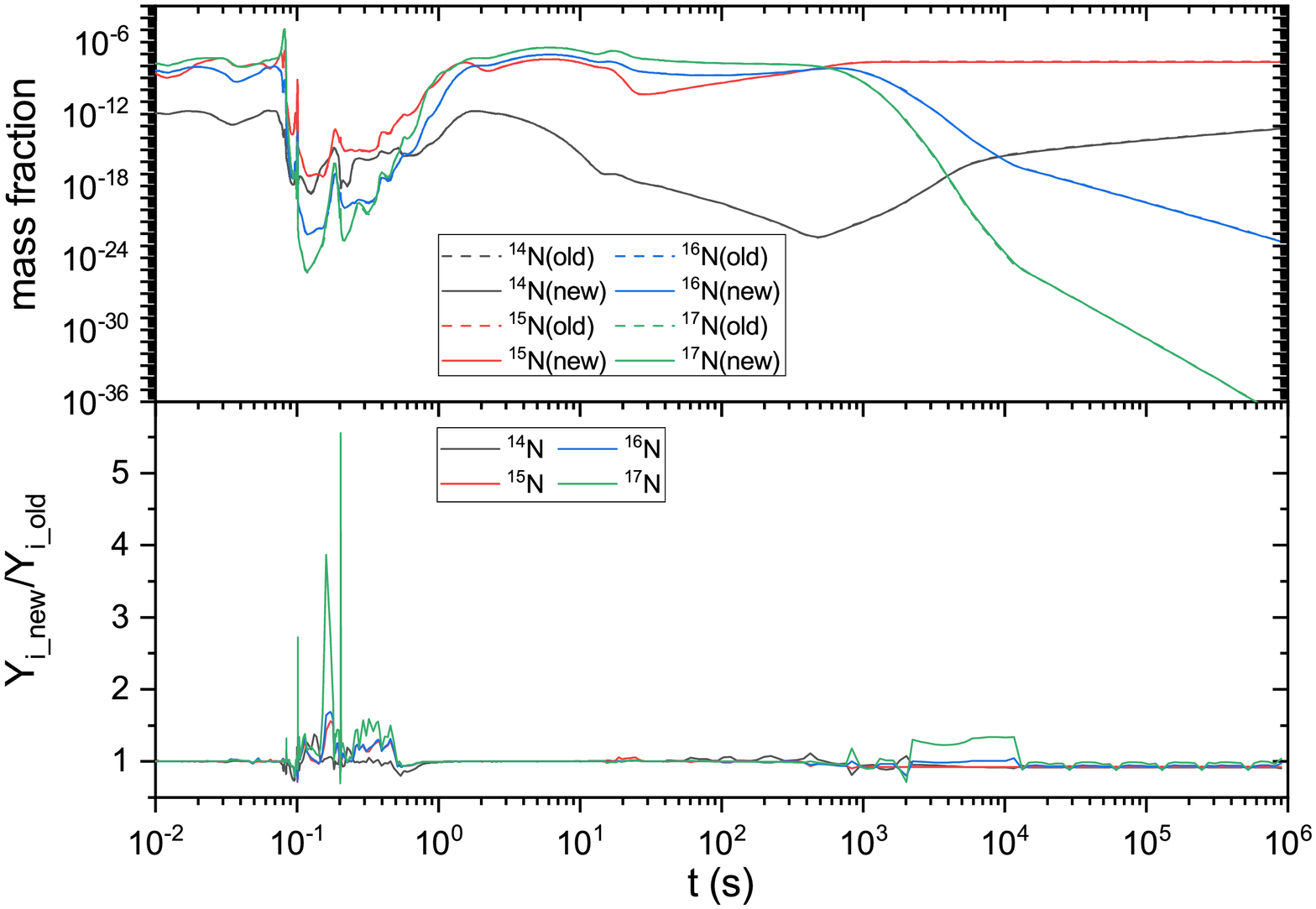}
\caption{\label{fig16} Time evolution of nitrogen isotopic abundances for the two $^{17}$O($n$,$\gamma$)$^{18}$O rates as in Fig.~\ref{fig15}.}
\end{center}
\end{figure}

\begin{figure}[t]
\begin{center}
\includegraphics[width=0.6\textwidth]{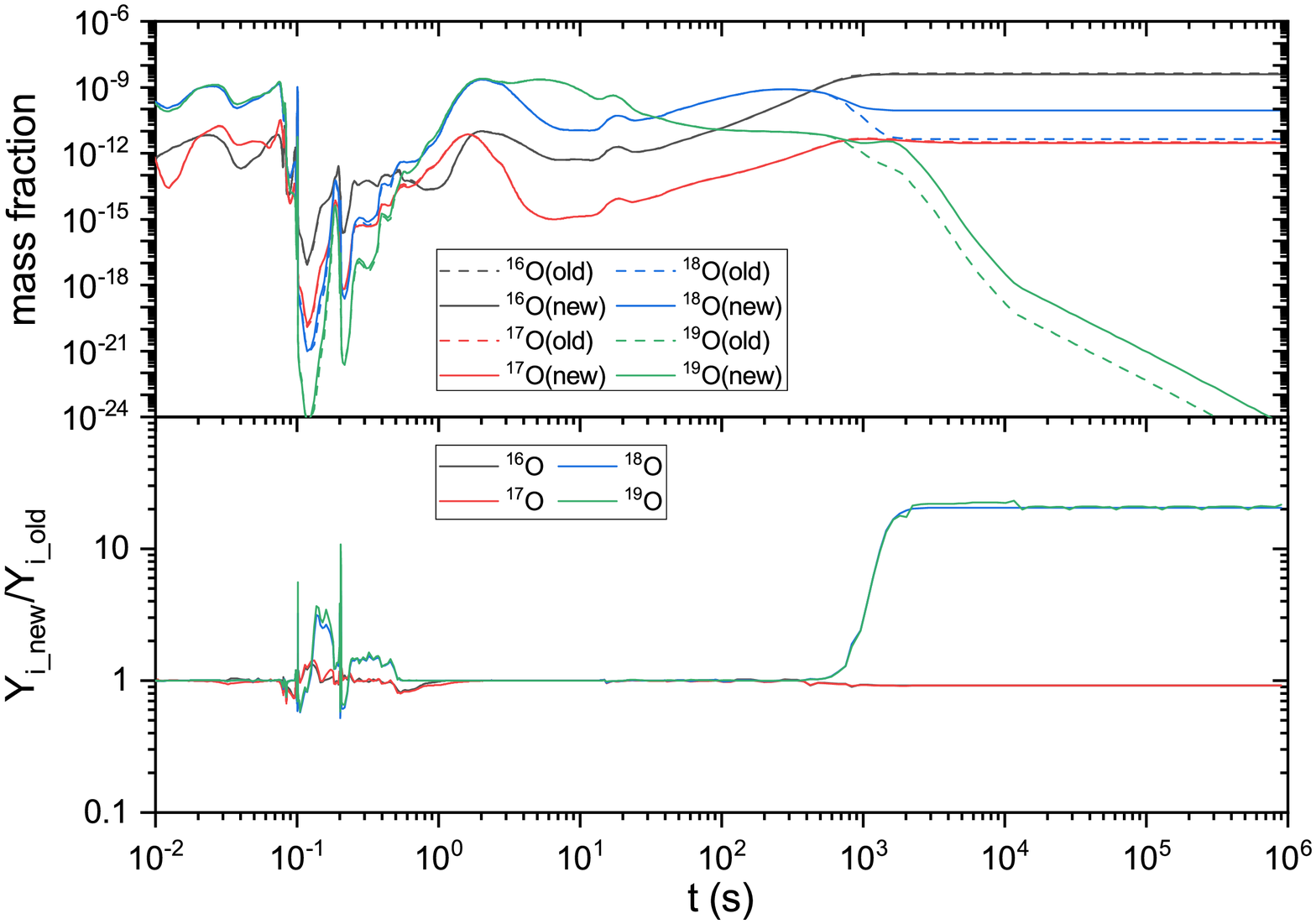}
\caption{\label{fig17} Time evolution of oxygen isotopic abundances for the two $^{17}$O($n$,$\gamma$)$^{18}$O rates as in Fig.~\ref{fig15}.}
\end{center}
\end{figure}

\begin{figure}[t]
\begin{center}
\includegraphics[width=0.6\textwidth]{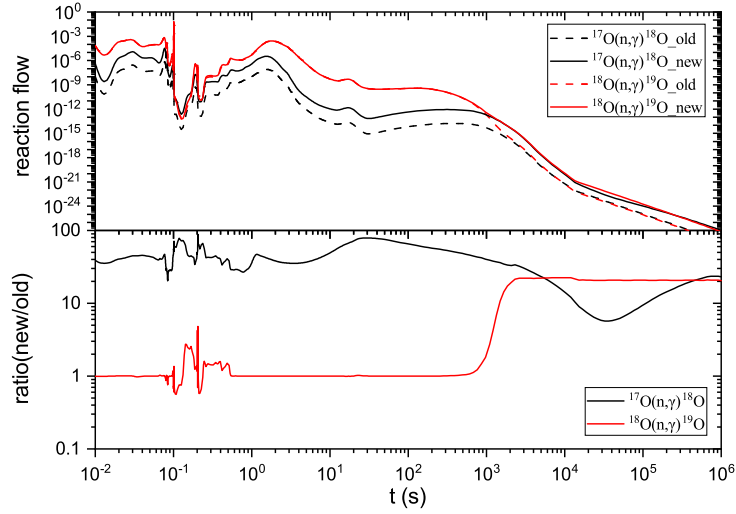}
\caption{\label{fig18} Time evolution of nuclear flows for $^{17}$O($n$,$\gamma$)$^{18}$O and $^{18}$O($n$,$\gamma$)$^{19}$O reactions in the $r$-process collapsar model
($Y_e$ = 0.05, without fission recycling effect). Here, old and new denote the calculations using the previous and present $^{17}$O($n$,$\gamma$)$^{18}$O rates, respectively.}
\end{center}
\end{figure}

\begin{figure}[t]
\begin{center}
\includegraphics[width=0.6\textwidth]{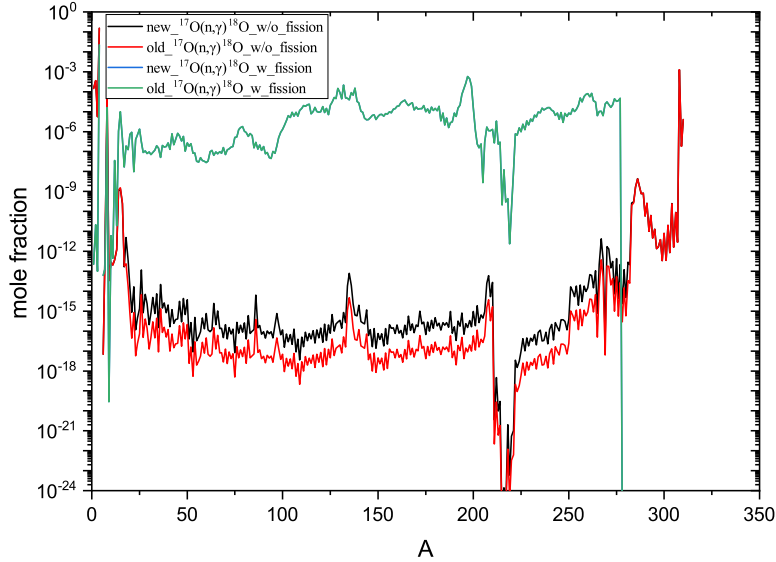}
\caption{\label{fig19} Final abundances of the $r$-process nucleosynthesis in the collapsar model calculated by using two $^{17}$O($n$,$\gamma$)$^{18}$O rates with and without
the fission recycling effect (with $Y_e$ = 0.05). The indexes, new and old, denote the calculated results by using the present and previous rates, respectively. The very small
difference in the fission recycling case is invisible in the figure.}
\end{center}
\end{figure}

\begin{figure}[t]
\begin{center}
\includegraphics[width=0.6\textwidth]{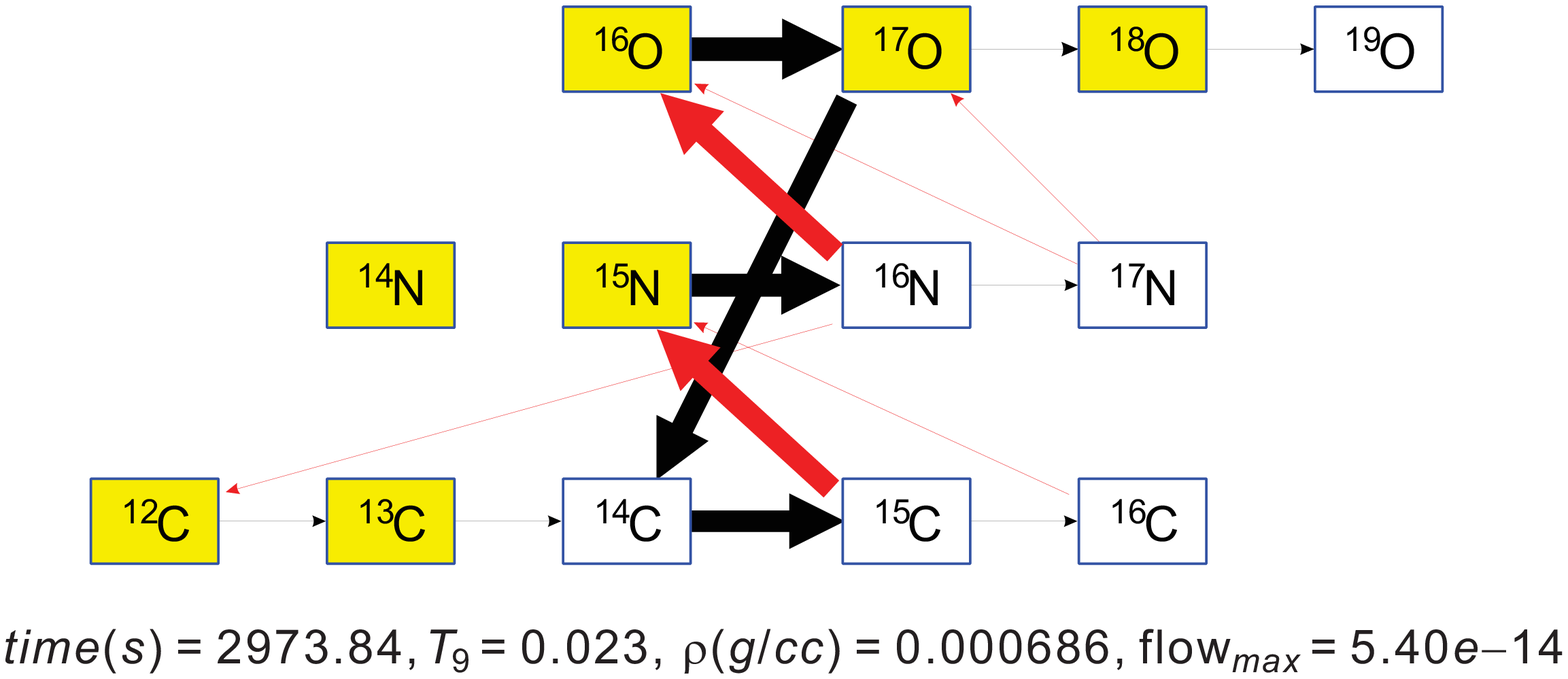}
\caption{\label{fig20} The nuclear flows in the collapsar $r$-process model at $t\approx$ 3000 s when the largest difference in $^{18}$O abundance is observed
($Y_e$ = 0.05, without fission recycling effect). The thickness of the arrow is proportional to the amplitude of the flow.}
\end{center}
\end{figure}

All nucleosynthetic products at earlier times $t<$ 0.1 s of collapsar event are once reset by the shock passage at $t\sim$ 0.1 s (see Figs. 8--11). At later times, the temperature is still
high enough ($T_9\sim$ 5) for charged-particle reactions to operate. Then, $\alpha$-induced reactions $^3$H($\alpha$,$\gamma$)$^7$Li and $^8$Li($\alpha$,$n$)$^{11}$B as well
as neutron capture flow play the important roles in the production of carbon, nitrogen, oxygen, and even heavier neutron-rich nuclei up to $A\sim$ 80~\citep{sas05}. Figures 9--11
show the time evolution of carbon, nitrogen, and oxygen isotopic abundances calculated with the present (solid lines) and previous (dashed lines) $^{17}$O($n$,$\gamma$)$^{18}$O
rates. Significant differences are seen between the two cases in the $^{15}$C, $^{17}$N and $^{18,19}$O abundances during the ``$\alpha$-process" within 0.25 s $<t<$ 1 s,
corresponding to a temperature region of 5 $>$ $T_9$ $>$ 2; also remarkable differences in $^{18,19}$O abundances are seen after the ``$\alpha$-process" at $t$ $>$~1000~sec,
corresponding to a temperature region of $T_9$ $<$~0.1. Compared to the abundances calculated with the previous $^{17}$O($n$,$\gamma$)$^{18}$O rate, the $^{18}$O and $^{19}$O
abundances calculated with the present rate are larger by almost the same factor of $\sim$20. This enhancement is made by the interplay between the strong neutron-capture flow
under the condition of extremely high neutron-number density and the $\alpha$-induced reactions in collapsar $r$-process, where the $\alpha$-particles are continuously provided by
the ($n$,$\alpha$) reactions during the $r$-process. Once $^3$H($\alpha$,$\gamma$)$^7$Li reaction produces $^7$Li, subsequent light nuclear reactions
$^7$Li($n$,$\gamma$)$^8$Li($\alpha$,$n$)$^{11}$B and $^8$Li($n$,$\gamma$)$^9$Li(e$^- \nu$)$^9$Be cross over the mass gap of unstable $A$ = 8 nuclear systems, and
radiative neutron-capture flow followed by the $\beta$-decay is established in the neutron-rich Be-B-C-N isotopes~\citep{ter01}. $^{18}$O is thus predominantly produced via the
$\beta$-decay of $^{18}$N in the neutron-capture flow at $t$ $<$~1000~s. Therefore, one cannot see any difference in $^{18}$O abundance until $t$ $\approx$~1000~s. However,
as the time passes by, neutrons are gradually exhausted and their number density rapidly decreases at around $t$ $\approx$~1000~s. Resultantly, the light nuclear reactions do not
provide neutron-rich Be-B-C-N isotopes any more. After this epoch of changing the environmental condition, $^{18}$O is now produced via
$^{16}$O($n$,$\gamma$)$^{17}$O($n$,$\gamma$)$^{18}$O reaction sequence (see Fig.~\ref{fig17}) by residual neutrons. Since the destruction rate of $^{18}$O via
$^{18}$O($n$,$\gamma$)$^{19}$O is unchanged, the produced $^{18}$O remains at higher abundance level in the case using larger cross section for $^{17}$O($n$,$\gamma$)$^{18}$O.
Accordingly, $^{19}$O abundance produced via $^{18}$O($n$,$\gamma$)$^{19}$O also stays at higher level by the same factor. Figure~\ref{fig19} shows the time evolution of nuclear
reaction flows through $^{17}$O($n$,$\gamma$)$^{18}$O and $^{18}$O($n$,$\gamma$)$^{19}$O, $|dY_i/dt|$ for $i$ = $^{17}$O and $^{18}$O, which supports this production
mechanism.

We have found an interesting ``loophole" effect in the present model. Compared to the $^{17}$O($n$,$\gamma$)$^{18}$O rate, the $^{17}$O($n$,$\alpha$)$^{14}$C rate is tremendously
larger (see Appendix B), and this results in a ``loop" of nuclear reaction network in which the abundance flows back to the lighter-mass nucleus $^{14}$C, yet taking away appreciable
abundance flow out of the $^{17}$O-$^{14}$C-$^{17}$O loop due to the enhanced new rate (or cross section) of the $^{17}$O($n$,$\gamma$)$^{18}$O reaction, named ``loophole" effect.
The enhancement of the $^{17}$O($n$,$\gamma$)$^{18}$O rate significantly changes the final nuclear abundances if fission recycling is not included in the model, as shown in
Fig.~\ref{fig19} (black and red solid lines). The present $^{17}$O($n$,$\gamma$)$^{18}$O rate results in an overall increase by one order of magnitude in the abundances of nuclei in
almost the whole mass region of 20~$<$ $A$ $<$~275. It can be explained by the fact that $^{17}$O($n$,$\gamma$)$^{18}$O works as a ``loophole" in the
$^{17}$O($n$,$\alpha$)$^{14}$C($n$,$\gamma$)$^{15}$C($\beta^-$)$^{15}$N($n$,$\gamma$)$^{16}$N($\beta^-$)$^{16}$O($n$,$\gamma$)$^{17}$O cycle (see Fig.~\ref{fig20} which
shows nuclear flows at $t\approx$ 3000~sec when the largest difference in $^{18}$O abundance is observed). Neutrons are exhausted through the ``loophole" mechanism without
producing new heavier nuclei. Although a competing $^{17}$O($n$,$\alpha$)$^{14}$C reaction takes a huge flux of abundance which is about two orders of magnitude larger than that
via the $^{17}$O($n$,$\gamma$)$^{18}$O reaction, the present larger $^{17}$O($n$,$\gamma$)$^{18}$O rate broadens a flow path to produce more seed nuclei for the subsequent
radiative neutron-capture process. Resultant increase in $^{18}$O and $^{19}$O abundances occur efficiently in accordance with the enhanced $r$-process nuclei when the temperature
of the system becomes as low as $T_9$ $<$~0.1 at $t$ $>$~300~sec. It is worthy of noting that although the maximum temperature can reach up to $T_9$ $\sim$~9 in the collapsar
model (see Fig.~\ref{fig14}), all the abundance enhancements illustrated here mainly emerge from the present $^{17}$O($n$,$\gamma$)$^{18}$O rate in the temperature region below
$T_9$ $\sim$~2.

We have calculated the $r$-process nucleosynthesis including a fission recycling based on the model of~\citet{fam20}, in which fission was implemented for the Cf isotopic chain, i.e.,
$^{270-311}$Cf, and the fission rates were assumed to be 100 s$^{-1}$ for all californium isotopes. As displayed in Fig.~\ref{fig19}, the calculated final abundances are almost identical
for the previous (green solid line) and present (blue solid line) $^{17}$O($n$,$\gamma$)$^{18}$O rates. The resulting difference is so small (less than 1\%) that it is invisible in the figure.
The final abundance is redistributed due to the fission recycling, especially in the mass region 100 $<A<$ 275. Therefore, the difference in the final abundances (at the order of magnitude
of 10$^{-15}$) observed in the case without fission is almost completely washed out by the fission recycling effect. The fission modes, i.e. spontaneous, $\beta$-delayed and neutron
capture-induced fissions, and the fission fragment distribution which were used in the present model are taken from the theoretical calculation used in~\citet{shi16}. Laboratory
experiments are still far to reach significant fissile region of such superheavy nuclei with large neutron-number excess~\citep{ahm21}. Further studies of the fission and the fission
recycling effects on collapsar $r$-process are highly desirable both experimentally and theoretically.

\begin{figure}[t]
\begin{center}
\includegraphics[width=0.6\textwidth]{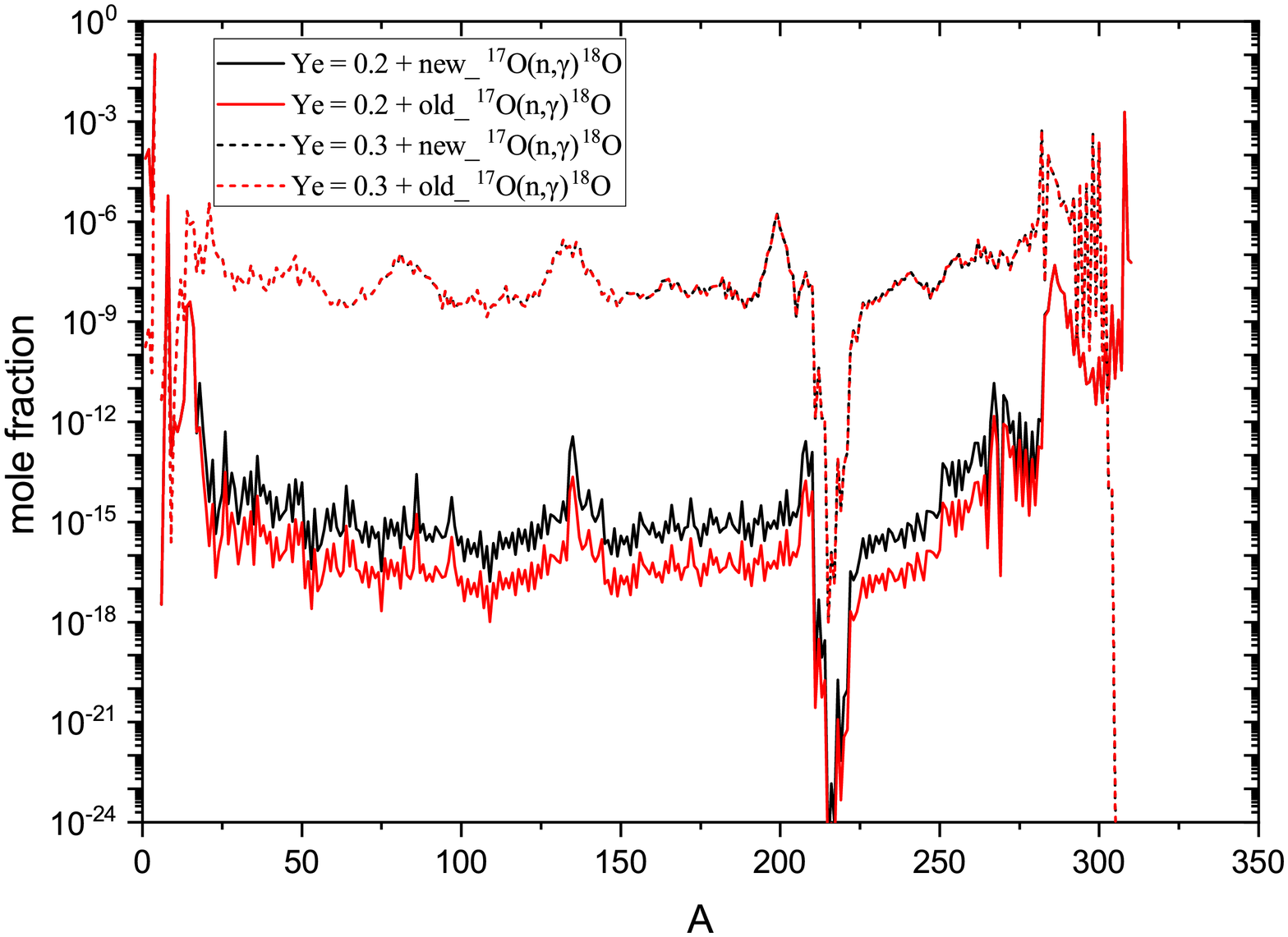}
\caption{\label{fig21} Final abundances of the $r$-process nucleosynthesis in the collapsar model calculated by using two $^{17}$O($n$,$\gamma$)$^{18}$O rates and
two different initial $Y_e$ values (without fission recycling effect).}
\end{center}
\end{figure}

\begin{figure}[t]
\begin{center}
\includegraphics[width=0.6\textwidth]{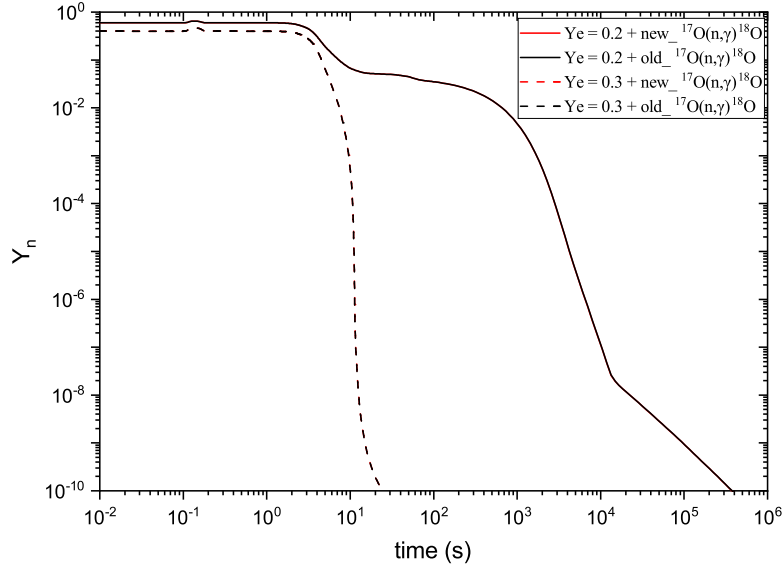}
\caption{\label{fig22} Time evolution of free neutron abundances in two different initial $Y_e$ cases in the collapsar $r$-process model (without fission recycling effect).
The differences between the two cases using new and old $^{17}$O($n$,$\gamma$)$^{18}$O rates are hardly distinguishable.}
\end{center}
\end{figure}

\begin{figure}[t]
\begin{center}
\includegraphics[width=0.6\textwidth]{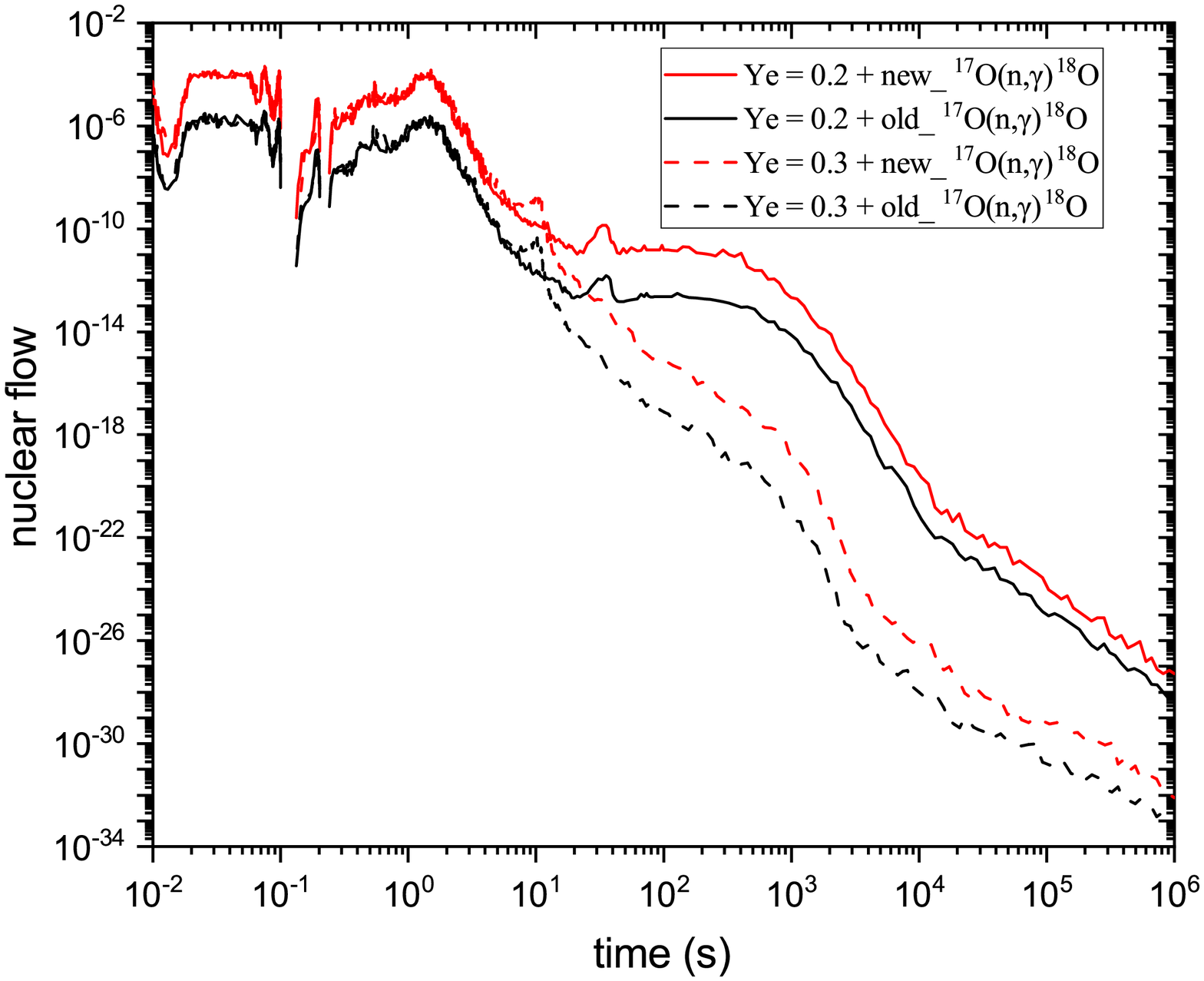}
\caption{\label{fig23} Time evolution of nuclear flows for $^{17}$O($n$,$\gamma$)$^{18}$O reaction in the collapsar model calculated by using two
$^{17}$O($n$,$\gamma$)$^{18}$O rates and two different initial $Y_e$ values (without fission recycling effect).}
\end{center}
\end{figure}

To study the influence of less neutron-rich conditions on the above $r$-process calculations, we have performed calculations with values of $Y_e$ = 0.1, 0.2, 0.3 and 0.4,
respectively. When fission is included in the network for either of the four $Y_e$ values, similarly to the above case with $Y_e$ = 0.05, the differences caused by the new
$^{17}$O($n$,$\gamma$)$^{18}$O rate are invisible in the final abundance pattern. However, the patterns vary between the cases by using the new and old rates if the fission recycling
is switched off artificially. As shown in Fig.~\ref{fig21}, the differences in abundances can be seen for $Y_e$ = 0.2 but not for $Y_e$ = 0.3. The results for $Y_e$ = 0.1 and 0.2, and those
for 0.3 and 0.4, are respectively analogous to each other. Therefore, we only focus on the cases with $Y_e$ = 0.2 and 0.3 as representative values in the following discussion.
Figure~\ref{fig22} shows the time evolution of free neutron abundances in the system with different initial $Y_e$ values and $^{17}$O($n$,$\gamma$)$^{18}$O rates. It shows that
$Y_e$ plays a more critical role in determining the evolution of neutron abundances than the reaction rate does. The abundant nuclei in the system after $\alpha$-process are those
having larger binding energies corresponding to the initial $Y_e$ value~\citep{mey94}.  Since $Y_e$ = 0.3 is closer to the proton-to-mass ratio $Z$/$A$ of the seed nucleus $^{78}$Ni for
the $r$-process, most neutrons are exhausted by the time when $^{78}$Ni is abundantly produced after the $\alpha$-process. As a result, the number of neutrons decreases drastically
after $t\sim$ 10 s, which suppresses the subsequent neutron-capture reactions. On the contrary, for the $Y_e$ = 0.2 case, plenty of free neutrons still remain in the system even at later
times and they could synthesize more abundant heavy nuclei in the region $A$ $>$ 305 as shown in Fig.~\ref{fig21}. The nuclear flow through the $^{17}$O($n$,$\gamma$)$^{18}$O
reaction, i.e., $\rho N_A\langle\sigma v\rangle Y_{\mathrm{^{17}O}}Y_n$, is obviously enhanced due to the larger new rate in both cases as displayed in Fig.~\ref{fig23}. However, this
enhancement in the $Y_e$ = 0.3 case is at the level of about four orders of magnitude smaller than that in the $Y_e$ = 0.2 case after $t\sim$ 10 s. Thus, for $Y_e$ = 0.3, the neutron
deficiency eliminates the influence of the new $^{17}$O($n$,$\gamma$)$^{18}$O rate on the final nucleosynthetic yield, as shown in Fig.~\ref{fig20}. Combined with the fact that nuclear
abundances in the mass region 20 $<A<$ 250 remain high in this case, the effect of new $^{17}$O($n$,$\gamma$)$^{18}$O rate is small.

\subsubsection{$r$-process in neutron bursts}
We also study the impact of new $^{17}$O($n$,$\gamma$)$^{18}$O reaction rate on a  neutron burst model~\citep{boj14}, which is developed for supernova nucleosynthesis of massive stars~\citep{rau02}. The model is based on the explosion of initially 15$M_\sun$ stellar model s15a28 for explosion energy $E$ = 1.0$\times 10^{51}$~erg~$\equiv$ 1.0 B~\citep{boj14}. Having confirmed that pressure waves in the post-shock material tend to homogenize the energy density dominated by the radiation~\citep{woo95}, the post-shock temperature $T$ and the other quantities are solved by the NucNet Tools $simple\_snII$~\citep{mey12}.

In order to find a most promising site where the new $^{17} $O($n$,$\gamma$)$^{18}$O rate could make the biggest effect on the $^{18}$O production, let us assume that the system is approximately in the steady state.
We consider here the following four possible reaction sequences as the main pathways for the production of $^{18}$O: $^{14}$N($\alpha$,$\gamma$)$^{18}$F($n$,$p$)$^{18}$O, $^{16}$O($n$,$\gamma$)$^{17}$O($n$,$\gamma$)$^{18}$O, $^{14}$N($n$,$p$)$^{14}$C($\alpha$,$\gamma$)$^{18}$O, $^{17}$O($n$,$\alpha$)$^{14}$C($\alpha$,$\gamma$)$^{18}$O.
We must include of course the competing destruction reactions of $^{18}$F and $^{14}$N: $^{18}$F($n$,$\alpha$)$^{15}$N and $^{14}$N($n$,$\alpha$)$^{17}$F.
We denote the $^{18}$F abundance upon this condition as $\tilde{Y}_{^{18}\mathrm{F}}$.
The first main reaction pathway $^{14}$N($\alpha$,$\gamma$)$^{18}$F($n$,$p$)$^{18}$O and $^{18}$F($n$,$\alpha$)$^{15}$N in the steady state leads to
\begin{eqnarray}
\rho N_\mathrm{A} Y_{^{14}\mathrm{N}} Y_{\alpha}\langle\sigma v\rangle_{^{14}\mathrm{N}(\alpha,\gamma)}=\rho N_\mathrm{A} Y_{\mathrm{n}} \tilde{Y}_{^{18}\mathrm{F}}\langle\sigma v\rangle_{^{18} \mathrm{F}(n,\alpha)}+\rho N_\mathrm{A} Y_{\mathrm{n}} \tilde{Y}_{^{18}\mathrm{F}}\langle\sigma v\rangle_{^{18} \mathrm{F}(n,p)}\, ,
\label{eq9}
\end{eqnarray}
and
$\tilde{Y}_{^{18}\mathrm{F}}$ is easily solved to be
\begin{eqnarray}
\tilde{Y}_{^{18}\mathrm{F}}=\frac{Y_{^{14}\mathrm{N}} Y_{\alpha}}{Y_{\mathrm{n}}} \frac{\langle\sigma v\rangle_{^{14}\mathrm{N}(\alpha,\gamma)}}{\langle\sigma v\rangle_{^{18}\mathrm{F}(n,\alpha)}+\langle\sigma v\rangle_{^{18}\mathrm{F}(n,p)}} .
\label{eq10}
\end{eqnarray}
Assuming that $^{17}$O and $^{14}$C are also in the steady state, their abundances can be solved similarly,
\begin{eqnarray}
\tilde{Y}_{^{17}\mathrm{O}}=Y_{^{16}\mathrm{O}} \frac{\langle\sigma v\rangle_{^{16}\mathrm{O}(n,\gamma)}}{\langle\sigma v\rangle_{^{17}\mathrm{O}(n,\alpha)}+\langle\sigma v\rangle_{^{17}\mathrm{O}(n,\gamma)}}
,
\label{eq11}
\end{eqnarray}

\begin{eqnarray}
\tilde{Y}_{^{14}\mathrm{C}}=Y_{\mathrm{n}} \frac{Y_{^{14}\mathrm{N}}\langle\sigma v\rangle_{^{14}\mathrm{N}(n,p)}+\tilde{Y}_{^{17}\mathrm{O}}\langle\sigma v\rangle_{^{17}\mathrm{O}(n,\alpha)}}{Y_{\alpha}\langle\sigma v\rangle_{^{14}\mathrm{C}(\alpha,\gamma)}} .
\label{eq12}
\end{eqnarray}
We now define time evolution of the yield of $^{18}$O in question through the reaction $A$($a$,$b$) as,
\begin{eqnarray}
\left.\frac{d Y_{^{18}\mathrm{O}}}{d t}\right|_{A(a,b)}=\rho N_\mathrm{A} Y_{A} Y_{a}\langle\sigma v\rangle_{A(a,b)} ,
\label{eq13}
\end{eqnarray}
in terms of the steady state abundances of $^{18}$F, $^{17}$O and $^{14}$C. Combining these Eqs.(10)-(13), we can determine time evolution of the yields of $^{18}$O through $^{18}$F($n$,$p$)$^{18}$O, $^{17}$O($n$,$\gamma$)$^{18}$O, and $^{14}$C($\alpha$,$\gamma$)$^{18}$O, respectively,
\begin{eqnarray}
\left.\frac{d Y_{^{18}\mathrm{O}}}{d t}\right|_{^{18}{\mathrm{F}(n,p)}}=\rho N_\mathrm{A} Y_{^{14}\mathrm{N}} Y_{\alpha} \frac{\langle\sigma v\rangle_{^{14}\mathrm{N}(\alpha,\gamma)}\cdot\langle\sigma v\rangle_{^{18} \mathrm{F}(n,p)}}{\langle\sigma v\rangle_{^{18}\mathrm{F}(n,\alpha)}+\langle\sigma v\rangle_{^{18}\mathrm{F}(n,p)}}
\label{eq14}
\end{eqnarray}

\begin{eqnarray}
\left.\frac{d Y_{^{18}\mathrm{O}}}{d t}\right|_{^{17}{\mathrm{O}(n,\gamma)}}=\rho N_\mathrm{A} Y_{^{16}\mathrm{O}} Y_{\mathrm{n}} \frac{\langle\sigma v\rangle_{^{16}\mathrm{O}(n,\gamma)}\cdot\langle\sigma v\rangle_{^{17} \mathrm{O}(n,\gamma)}}{\langle\sigma v\rangle_{^{17}\mathrm{O}(n,\alpha)}+\langle\sigma v\rangle_{^{17}\mathrm{O}(n,\gamma)}}
\label{eq15}
\end{eqnarray}

\begin{eqnarray}
\left.\frac{d Y_{^{18}\mathrm{O}}}{d t}\right|_{^{14}{\mathrm{C}(\alpha,\gamma)}}=\rho N_\mathrm{A} Y_{\mathrm{n}}\left(Y_{^{14}\mathrm{N}}\langle\sigma v\rangle_{^{14}\mathrm{N}(n,p)} +Y_{^{16}\mathrm{O}}\frac{\langle\sigma v\rangle_{^{16}\mathrm{O}(n,\gamma)}\cdot\langle\sigma v\rangle_{^{17} \mathrm{O}(n,\alpha)}}{\langle\sigma v\rangle_{^{17}\mathrm{O}(n,\alpha)}+\langle\sigma v\rangle_{^{17}\mathrm{O}(n,\gamma)}}\right).
\label{eq16}
\end{eqnarray}

\begin{figure}[t]
\begin{center}
\includegraphics[width=0.6\textwidth]{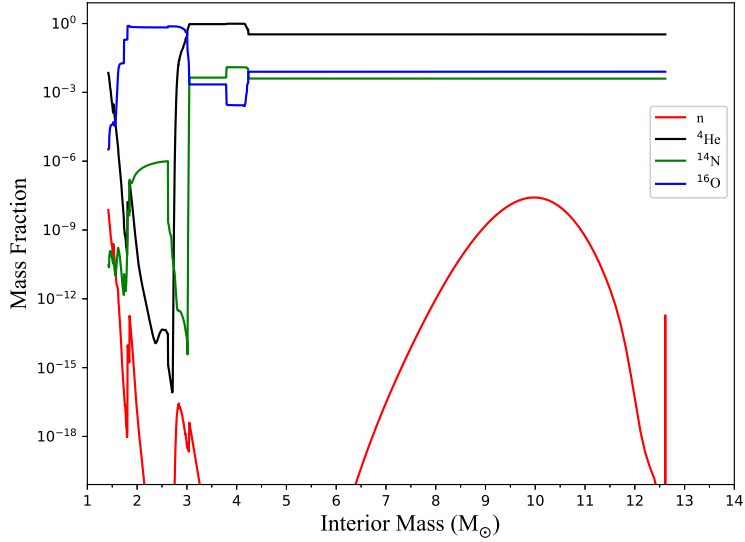}
\caption{\label{fig24} Mass fraction of some key species as a function of Lagrangian coordinate $M_r$ after the $E$ = 1.0 B explosion of s15a28.}
\end{center}
\end{figure}

\noindent
From these equations, we can find a most promising site where the distinct difference in the $^{18}$O abundance could arise from our larger $^{17}$O($n$,$\gamma$)$^{18}$O rate. It is obvious that such an ideal condition should satisfy the higher initial abundances of neutron and $^{16}$O than those of $^{4}$He and $^{14}$N so that the reaction sequence
$^{16}$O($n$,$\gamma$)$^{17}$O($n$,$\gamma$)$^{18}$O makes predominant contribution to the production of $^{18}$O. Zone 564 at $M_r$ = 2.82~$M_{\sun}$ and initial temperature $T_9$ = 1.12 in Fig.~\ref{fig24} is thus identified to be a most promising site to see the effect of new $^{17}$O($n$,$\gamma$)$^{18}$O rate.

In the above discussions, we confine ourselves to several limited number of most effective nuclear reactions on the production of $^{18}$O. However, once we identify the zone 564, the $r$-process nucleosynthesis calculation is performed by using exactly the same full nuclear reaction network adopted in collapsar r-process as discussed in the previous subsection.

\begin{figure}[t]
\begin{center}
\includegraphics[width=0.6\textwidth]{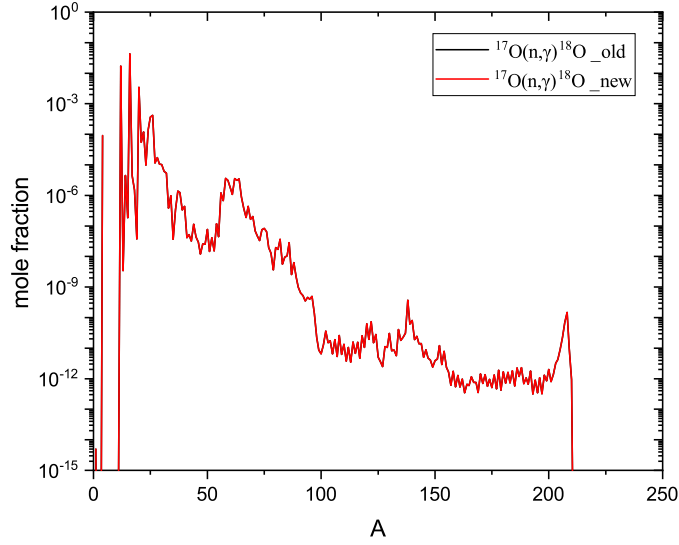}
\caption{\label{fig25} Final abundances of the $r$-process nucleosynthesis with fission recycling switched off in zone 564 at 10$^6$ seconds after the $E$ = 1.0 B explosion of s15a28 by using two $^{17}$O($n$,$\gamma$)$^{18}$O rates.}
\end{center}
\end{figure}

\begin{figure}[t]
\begin{center}
\includegraphics[width=0.6\textwidth]{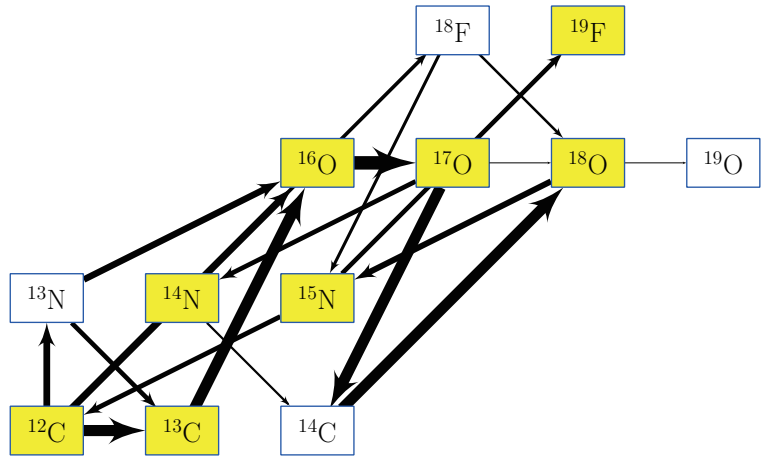}
\caption{\label{fig26} Integrated current diagram at 10$^6$ seconds after the $E$ = 1.0 B explosion of s15a28 using the new $^{17}$O($n$,$\gamma$)$^{18}$O rate.}
\end{center}
\end{figure}

The calculated final abundances in this zone are shown in Fig.~\ref{fig25}, where the impact of the new $^{17}$O($n$,$\gamma$)$^{18}$O rate is hardly seen. The final abundances of the key isotopes of carbon, nitrogen, oxygen, and fluorine are listed in Table~\ref{tab3_key_species}. The yield of $^{18}$O is enhanced by only 0.2\% due to the new larger $^{17}$O($n$,$\gamma$)$^{18}$O rate, and less than 1\% difference is found in the yields of the other isotopes.
This result is explained by defining the integrated current for the reaction $r$ that links reactant $j$ and product $i$ as
\begin{eqnarray}
I[r]\left(t, t_{0}\right)=\int_{t_{0}}^{t}\left[\left(\frac{d Y_{i}}{d t}\right)_{r: j \rightarrow i}-\left(\frac{d Y_{i}}{d t}\right)_{r: i \rightarrow j}\right] d t^{\prime}.
\label{eq17}
\end{eqnarray}
\noindent
Fig.~\ref{fig26} displays the significance of nuclear reactions where the thickness of each arrow is proportional to the magnitude of the integrated current for that reaction ${r: j \rightarrow i}$.
It is indicated that the integrated current for a competing reaction $^{17}$O($n$,$\alpha$)$^{14}$C takes about two orders of magnitude larger current to produce $^{14}$C than the current for $^{17}$O($n$,$\gamma$)$^{18}$O. Subsequently, the condition of high temperature and sufficiently abundant alpha particles facilitates $^{14}$C to capture $\alpha$ particles. Even in the ideal zone 564 which is most favorable for the production of $^{18}$O through $^{16}$O($n$,$\gamma$)$^{17}$O($n$,$\gamma$)$^{18}$O, the $^{14}$C($\alpha$,$\gamma$)$^{18}$O reaction makes the largest contribution to the yield of $^{18}$O, thus the change of $^{17}$O($n$,$\gamma$)$^{18}$O rate is highly suppressed. We conclude that impact of the present new larger $^{17}$O($n$,$\gamma$)$^{18}$O rate on the neutron burst model is marginal.

\begin{table*}
\footnotesize
\centering
\caption{\label{tab3_key_species} Final abundances of key species at 10$^{6}$ seconds after the $E$ = 1.0 B explosion of s15a28 by using two $^{17}$O($n$,$\gamma$)$^{18}$O rates.}
\begin{tabular}{ccccc}
\hline
  & $^{12}$C  & $^{13}$C  & $^{14}$N & $^{15}$N \\
\hline
new rate & 1.7104E-02 & 3.4205E-09 & 7.3949E-08 & 1.8553E-07 \\
old rate & 1.7104E-02 & 3.4209E-09 & 7.4000E-08 & 1.8543E-07 \\
ratio(new/old) & 1.0000E+00 & 9.9988E-01 & 9.9931E-01 & 1.0005E+00 \\
\hline
  & $^{16}$C  & $^{17}$O  & $^{18}$O & $^{19}$F \\
\hline
new rate & 4.3387E-02 & 4.3715E-06 & 1.5002E-06 & 3.7045E-08 \\
old rate & 4.3387E-02 & 4.3721E-06 & 1.4999E-06 & 3.7024E-08 \\
ratio(new/old) & 1.0000E+00 & 9.9985E-01 & 1.0002E+00 & 1.0006E+00 \\
\hline
\end{tabular}
\end{table*}

\section{Summary and outlook}
\label{sec5}
In this work, we have obtained a new thermonuclear $^{17}$O($n$,$\gamma$)$^{18}$O rate based on theoretical calculation of the DC as well as the resonance contributions, for a
temperature up to $T_9$ = 2 of general astrophysical interest. The associated uncertainties of this rate are also estimated based on reasonable parameter variations in the Monte-Carlo
calculations. It shows that the present rate is larger than that adopted in the JINA REACLIB library in the temperature region of $T_9$ = 0.01 $\sim$ 2 significantly, by up to a factor of
$\sim$80. In addition, for completeness, the $^{17}$O($n$,$\gamma$)$^{18}$O rate in the high temperature region of $T_9$ = 2 $\sim$ 10 has been obtained by normalizing the ENDF/B-VIII
data in the energy region of 1.0--20 MeV to the presently calculated cross section data around 1 MeV.

The impacts of the change in the $^{17}$O($n$,$\gamma$)$^{18}$O rate on the evolution of light element abundances in AGB stars and massive stars have been investigated in both main and weak $s$-process models. In the main $s$-process model for metal deficient stars, $^{18}$O and $^{19}$F abundances are found to be enhanced by factors of $\sim 40$ and $\sim 20$ in LP625-44 ([Fe/H] $= -2.70$) and CS31062-012/LP706-7 ([Fe/H] $= -2.58$), respectively, due to new larger $^{17}$O($n$,$\gamma$)$^{18}$O rate, still reproducing the observed abundances of heavy $s$-process elements fairly well. The new rate also results in ${\mathcal O}(1)$ \% modifications of neutron exposure under metal-poor conditions.
However, as for the weak $s$-process in massive stars, there is no discernable difference in nuclear abundances when a result for the new rate is compared with that of the JINA rate for both of core He and shell C burning stages.
The new rate of the $^{17}$O($n$,$\gamma$)$^{18}$O reaction results in $^{18}$O production via the reaction much more efficient than the old rate does. However,
because the $\beta$-decay of $^{18}$F and $^{14}$C($\alpha$,$\gamma$)$^{18}$O reaction are predominant in the $^{18}$O production, the sensitivity of $^{18}$O abundance to the $^{17}$O($n$,$\gamma$)$^{18}$O reaction rate is still very small. Furthermore, the impact of the present $^{17}$O($n$,$\gamma$)$^{18}$O rate on the $r$-process nucleosynthesis
in the collapsar model has been investigated. It shows that when the fission recycling is not included, the $^{18}$O and $^{19}$O abundances calculated with the present rate are larger
by almost a factor of $\sim$20 during and after the ``$\alpha$-process" phase, compared to those calculated with the previous JINA rate. An interesting ``loophole" effect is found
owing to the enhanced new $^{17}$O($n$,$\gamma$)$^{18}$O rate, which may significantly change the final nuclear abundances if fission recycling is not included in the model.
In this case, an overall increase by one order of magnitude was observed in the abundances of nuclei in almost the whole mass region of 20 $<A<$ 275. However, such significant
differences are almost completely washed out by the fission recycling effect, with which the final abundances are redistributed. The effect of new $^{17}$O($n$,$\gamma$)$^{18}$O rate on $r$-process in neutron burst model is also studied. In this model, the condition of high temperature and sufficiently abundant alpha particles makes $^{17}$O(n,$\alpha$)$^{14}$C($\alpha$,$\gamma$)$^{18}$O much more efficient than $^{17}$O($n$,$\gamma$)$^{18}$O
for the production of $^{18}$O. It shows that $^{14}$C($\alpha$,$\gamma$)$^{18}$O contributes overwhelmingly to the yield of $^{18}$O, and the impact of new $^{17}$O($n$,$\gamma$)$^{18}$O rate is highly suppressed.

So far, there are no experimental data available for the $^{17}$O($n$,$\gamma$)$^{18}$O reaction except for the thermal $\sigma_\mathrm{th}$ value. Therefore, the present calculated
cross sections above the $E_n$ = 10$^{-5}$ MeV energy region need further experimental clarification. Besides the direct neutron beam experiments, the time-reversal photo-dissociation
of $^{18}$O experiments may be also an option. In addition, the $^{14}$C($\alpha$,$\gamma$)$^{18}$O reaction rate still bears very large uncertainties in the AGB temperature regime
(up to a factor of $\sim$10;~\citet{ili10}), and hence the precise measurement of this reaction is strongly required in the low energy region, to better constrain the $^{18}$O abundance
in the $s$-process models.

\acknowledgments

This work was financially supported by the National Natural Science Foundation of China (Nos. 11825504, 12075027, 11850410441, 11961141004),
and by Grants-in-Aid for Scientific Research of Japan Society for the Promotion of Science (Nos. 20K03958, 17K05459).

\software{RADCAP~\citep{ber03}, MESA~\citep{pax11}, NucNet Tools~\citep{mey12}}

\appendix

\begin{figure}[tbp]
\begin{center}
\includegraphics[width=0.6\textwidth]{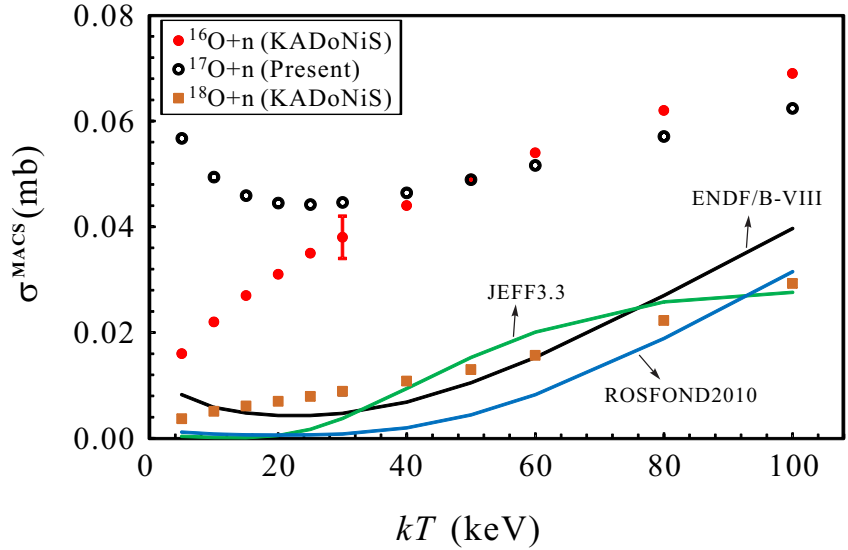}
\caption{\label{figapp1} Comparison of MACSs for $^{17}$O($n$,$\gamma$), $^{16}$O($n$,$\gamma$) and $^{18}$O($n$,$\gamma$). The MACSs for the latter two reactions are taken
from KADoNiS. In addition, the calculated MACSs for the $^{17}$O($n$,$\gamma$)$^{18}$O reaction by using the ENDF/B-VIII, JEFF3.3 and ROSFOND2010 libraries are indicated by
the solid lines, respectively. See text for details.}
\end{center}
\end{figure}

\begin{figure}[tbp]
\begin{center}
\includegraphics[width=0.6\textwidth]{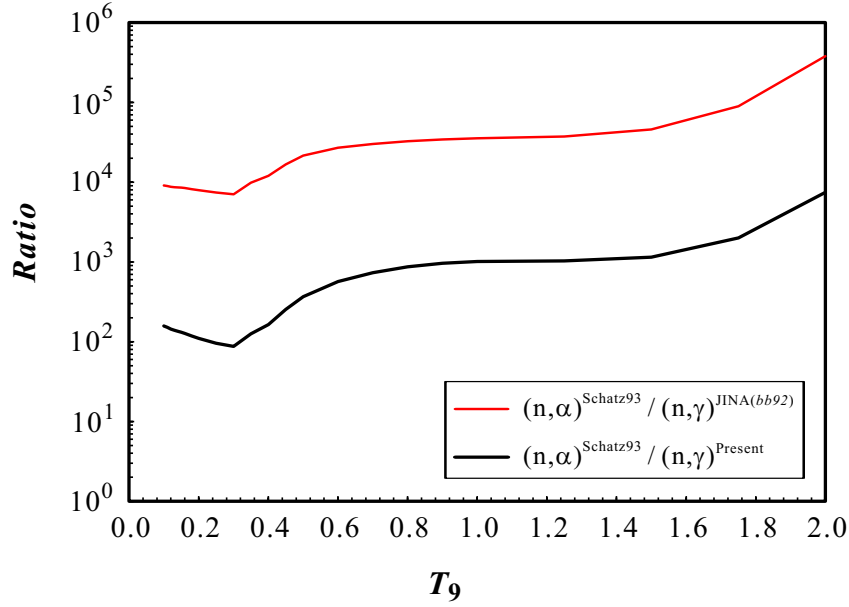}
\caption{\label{figapp2} Ratios of $^{17}$O($n$,$\alpha$)$^{14}$C and $^{17}$O($n$,$\gamma$)$^{18}$O rates. The label ``Schatz93'' indicates the~\citet{sch93} rate, and
``JINA (bb92)" indicates the \emph{bb92} rate compiled by JINA REACLIB.}
\end{center}
\end{figure}

\section{MACS of $^{17}$O($\lowercase{n}$,$\gamma$)$^{18}$O}
The Maxwellian-averaged cross sections (MACSs) of $^{17}$O($n$,$\gamma$)$^{18}$O are calculated by using the presently calculated cross sections, as well as the evaluated data
compiled in the recent ENDF/B-VIII, JEFF3.3 and ROSFOND2010 libraries. The resulting MACSs are shown in Fig.~\ref{figapp1}. Also the MACSs for $^{16}$O($n$,$\gamma$) and
$^{18}$O($n$,$\gamma$) which are compiled in the KADoNiS\footnote{Karlsruhe Astrophysical Database of Nucleosynthesis in Stars (KADoNiS), https://www.kadonis.org/}
~\citep{bao00,dil06,dil09}, are shown for comparison. It shows that the present MACSs of $^{17}$O($n$,$\gamma$) are close to those of $^{16}$O($n$,$\gamma$) in the temperature
region of $kT>$ 40 keV, below which they are quite different. The $^{17}$O($n$,$\gamma$) MACSs are significantly larger than those of $^{18}$O($n$,$\gamma$)$^{19}$O in the whole
temperature region, by a factor of 2--16. In addition, the presently calculated MACSs based on the evaluated libraries (indicated by three solid lines) are close to those of
$^{18}$O($n$,$\gamma$)$^{19}$O. In Section 1, we mentioned that \citet{nis09} assumed the $^{17}$O($n$,$\gamma$)$^{18}$O rate equaled to that of
$^{16}$O($n$,$\gamma$)$^{17}$O in the temperature region of AGB stars, and here we think this assumption is basically reasonable. But the predictions made by \cite{yam10} can be
constrained better with the present more reliable rate.

\section{$^{17}$O($\lowercase{n}$,$\alpha$)$^{14}$C rate}
For the $^{17}$O($n$,$\alpha$)$^{14}$C reaction, \citet{koe91} measured its cross section from thermal energy to approximately 1 MeV and derived a reaction rate in a temperature
region up to $T_9$ = 5, together with the experimental data of $^{14}$C($\alpha$,$n$)$^{17}$O reaction measured by~\citet{san56}. Later on, \citet{sch93} measured
the $^{17}$O($n$,$\alpha$)$^{14}$C cross section in the neutron energy range from 10 to 250 keV, and corrected the rate of~\citet{koe91} by about 10\% owing to the anisotropy effect.
Here, we have fitted two rates graphically shown in the previous work in the standard format of~\citet{rau00} as shown below.
I) The rate of~\citet{koe91} shown in their Fig. 5 (labelled as a solid line) can be fitted by,
\begin{eqnarray}
N_A\langle\sigma v\rangle_{(n,\alpha)} &=& \mathrm{exp}(27.9671+\frac{0.00030507}{T_9}+\frac{0.29888}{T_9^{1/3}}-24.2927T_9^{1/3}+25.4217T_9-34.5487T_9^{5/3}+2.55092\ln{T_9}) \nonumber \\
                          &+& \mathrm{exp}(72.2274-\frac{0.055398}{T_9}+\frac{18.496}{T_9^{1/3}}-80.6603T_9^{1/3}+6.9012T_9-0.57146T_9^{5/3}+28.1568\ln{T_9}) \nonumber \\
                          & &  \
\label{eq9}
\end{eqnarray}
with a fitting error of less than 1\% over the temperature region of $T_9$ = 0.001 $\sim$ 5;
II) The rate of~\citet{sch93}, which was shown as a solid line in their Fig.~6, can be fitted by,
\begin{eqnarray}
N_A\langle\sigma v\rangle_{(n,\alpha)} &=& \mathrm{exp}(-2145.6-\frac{100.742}{T_9}+\frac{1904.1}{T_9^{1/3}}+683.59T_9^{1/3}-400.627T_9+75.8261T_9^{5/3}+581.563\ln{T_9}) \nonumber \\
                          &+& \mathrm{exp}(3038.36+\frac{46.9075}{T_9}-\frac{2990.02}{T_9^{1/3}}-3039.63T_9^{1/3}+5760.5T_9-3140.75T_9^{5/3}-1672.1\ln{T_9}) \nonumber \\
                          &+& \mathrm{exp}(1029.25+\frac{9.41958}{T_9}-\frac{1003.42}{T_9^{1/3}}-1351.56T_9^{1/3}+2794.69T_9-1732.65T_9^{5/3}-622.95\ln{T_9}) \nonumber \\
                          & &  \
\label{eq10}
\end{eqnarray}
with a fitting error of less than 2\% over the temperature region of $T_9$ = 0.1 $\sim$ 2.

In fact, the discrepancies between the available experimental ($n$,$\alpha$) cross section data~\citep{koe91,sch93,san56} are still significant in the energy region of
$E_n \sim$ 40--600 keV, and the resulting rates can be very different, by up to a factor of 2. Therefore, further experimental verification is needed to clarify such discrepancies.
In addition, it should be noted that there are two $^{17}$O($n$,$\alpha$)$^{14}$C rates (i.e., \emph{kg91} and \emph{bb92}) collected in the JINA REACLIB. The \emph{kg91} rate is
actually the analytic one expressed by Eq.~4 of~\citet{koe91}, which was simply based on the thermal $\sigma_\mathrm{th}$($n$,$\alpha$) value and the parameters for the
$E_\mathrm{R}$ = 169, 238 keV resonances~\citep{wei58}. This analytic rate deviated greatly from the numerical results based on the experimental ($n$,$\alpha$) data, as shown in
their Fig.~5. Therefore, this analytic rate should not be utilized in the nucleosynthesis models, and it should be removed from the library for safety. In JINA REACLIB, the source of \emph{bb92}
rate was cited as~\citet{rau94}, who did not give an analytic equation for the $^{17}$O($n$,$\alpha$)$^{14}$C reaction rate. We found that the \emph{bb92} rate deviates from
the~\citet{koe91} numerical one up to $\sim$20\%, and hence the \emph{bb92} rate should be replaced by the well fitted rate with Eq.~\ref{eq9}, which is actually used in the present
$r$-process model. In addition, we suggest the rate of~\citet{sch93}, which is well fitted by Eq.~\ref{eq10}, should be compiled into the library in the future (but the temperature range
is limited to $T_9$ $\leq$~2).

For the $^{17}$O($n$,$\gamma$)$^{18}$O and $^{17}$O($n$,$\alpha$)$^{14}$C reactions, their corresponding reaction rate ratios are shown in Fig.~\ref{figapp2}. Owing to the present
large ($n$,$\gamma$) rate, the ($n$,$\alpha$)/($n$,$\gamma$) ratio reduced by factors of 35 $\sim$ 80, and the minimal ratio reduced to $\sim$90 from the previous value of
$\sim$7000 around $T_9$ = 0.3.

\end{document}